\documentclass[fleqn,usenatbib]{mnras}

\usepackage{newtxtext,newtxmath}
\usepackage[T1]{fontenc}

\DeclareRobustCommand{\VAN}[3]{#2}
\let\VANthebibliography\thebibliography
\def\thebibliography{\DeclareRobustCommand{\VAN}[3]{##3}\VANthebibliography}

\usepackage{graphicx}	
\usepackage{amsmath}	
\usepackage{orcidlink}

\usepackage{ulem}
\usepackage{xcolor}

\usepackage{subfigure}
\usepackage{multirow}
\usepackage{rotating}
\usepackage{tabularx}
\usepackage{adjustbox}
\usepackage{pdflscape}
\usepackage{booktabs}  
\usepackage{outlines}

\usepackage[utf8]{inputenc}
\usepackage[T1]{fontenc}
\usepackage{amsmath, amsfonts}
\usepackage{enumitem}

\def\msun{M$_{\odot}$}
\newcommand{\code}[1]{\texttt{#1}}
\def\v1d{{\code{V1D}}}

\def\cmfgen{{\code{CMFGEN}}}

\def\one{{\,\sc i}}
\def\two{{\,\sc ii}}
\def\three{{\,\sc iii}}
\def\four{{\,\sc iv}}
\def\five{{\sc v}}
\def\six{{\sc vi}}

\newcommand{\atlas}{\textit{\rm ATLAS}}

\newcommand{\kms}{\textrm{km\,s$^{-1}$}}

\newcommand{\ergs}{\textrm{erg\,s$^{-1}$}}

\newcommand{\msolar}{M$_{\odot}$}
\newcommand{\ml}{M$_{\odot}$\,yr$^{-1}$}

\title{A Low-Mass Helium Star Progenitor Model for the Type Ibn SN 2020nxt}
\author[Q.~Wang et al]{Qinan Wang\orcidlink{0000-0003-2238-1572}$^{1}$, Anika Goel$^{2,3}$, Luc Dessart\orcidlink{0000-0003-0599-8407}$^{4}$,
Ori D. Fox\orcidlink{0000-0003-2238-1572}$^{2}$, Melissa Shahbandeh\orcidlink{0000-0002-9301-5302}$^{1, 2}$, Sofia Rest$^{5}$, 
\newauthor Armin Rest\orcidlink{0000-0002-4410-5387}$^{1,2}$, Jose~H. Groh$^{6}$, Andrew Allan\orcidlink{0000-0002-3900-5111}$^{6}$, Claes Fransson\orcidlink{0000-0001-8532-3594}$^{7}$, Nathan Smith\orcidlink{0000-0001-5510-2424}$^{8}$, Griffin Hosseinzadeh\orcidlink{0000-0002-0832-2974}$^{8}$, \newauthor
Alexei V. Filippenko\orcidlink{0000-0003-3460-0103}$^{9}$, Jennifer Andrews\orcidlink{0000-0003-0123-0062}$^{10}$,  K. Azalee Bostroem\orcidlink{0000-0002-4924-444X}$^{8}$\thanks{LSSTC Catalyst Fellow}, Thomas G. Brink\orcidlink{0000-0001-5955-2502}$^{9,11}$, 
Peter Brown\orcidlink{0000-0001-6272-5507}$^{12,13}$, \newauthor Jamison Burke\orcidlink{0000-0003-0035-6659}$^{14,15}$, Roger Chevalier\orcidlink{0000-0002-9117-7244}$^{16}$,
Geoffrey C. Clayton\orcidlink{0000-0002-0141-7436}$^{17}$,
 Mi Dai\orcidlink{0000-0002-5995-9692}$^{1}$, Kyle W. Davis\orcidlink{0000-0002-5680-4660}$^{18}$, \newauthor Ryan~J.~Foley\orcidlink{0000-0002-2445-5275}$^{18}$, 
 Sebastian~Gomez\orcidlink{0000-0001-6395-6702}$^{1}$, Chelsea Harris\orcidlink{0000-0002-1751-7474}$^{19}$, Daichi Hiramatsu\orcidlink{0000-0002-1125-9187}$^{20,21}$, D. Andrew Howell\orcidlink{0000-0003-4253-656X}$^{14,15}$,
\newauthor  Connor~Jennings$^{9}$, Saurabh W.~Jha\orcidlink{0000-0001-8738-6011}$^{22}$,  Mansi M. Kasliwal\orcidlink{0000-0002-5619-4938}$^{23}$, Patrick L. Kelly\orcidlink{0000-0003-3142-997X}$^{24}$, Erik C. Kool\orcidlink{0000-0002-7252-3877}$^{7}$,
\newauthor Evelyn~Liu$^{9}$, Emily~Ma$^{9}$, Curtis McCully\orcidlink{0000-0001-5807-7893}$^{14}$, Adam M. Miller\orcidlink{0000-0001-9515-478X}$^{25,26}$, Yukei Murakami\orcidlink{0000-0002-8342-3804}$^{1}$, 
\newauthor Craig Pellegrino\orcidlink{0000-0002-7472-1279}$^{14,15}$, Estefania Padilla Gonzalez\orcidlink{0000-0003-0209-9246}$^{14,15}$, Derek~Perera$^{9}$, Justin Pierel\orcidlink{0000-0002-2361-7201}$^{2}$, César Rojas-Bravo\orcidlink{0000-0002-7559-315X}$^{18}$, 
\newauthor Matthew~R.~Siebert\orcidlink{0000-0003-2445-3891}$^{2}$, Jesper~Sollerman\orcidlink{0000-0003-1546-6615}$^{7}$, Tam\'as Szalai\orcidlink{0000-0003-4610-1117}$^{27,28}$, Samaporn~Tinyanont\orcidlink{0000-0002-1481-4676}$^{29}$, 
\newauthor Schuyler D. Van Dyk\orcidlink{0000-0001-9038-9950}$^{30}$, WeiKang Zheng\orcidlink{0000-0002-2636-6508}$^{9,31}$, Kenneth~C.~Chambers\orcidlink{0000-0001-6965-7789}$^{32}$, David~A.~Coulter\orcidlink{0000-0003-4263-2228}$^{18}$, 
\newauthor Thomas de Boer\orcidlink{0000-0001-5486-2747}$^{32}$, Nicholas~Earl\orcidlink{0000-0003-1714-7415}$^{33}$, Diego~Farias\orcidlink{0000-0002-6886-269X}$^{34}$,  Christa Gall\orcidlink{0000-0002-8526-3963}$^{34}$, Peter McGill\orcidlink{0000-0002-1052-6749}$^{18}$, 
\newauthor Conor~L.~Ransome\orcidlink{0000-0003-4175-4960}$^{35}$, Kirsty~Taggart\orcidlink{0000-0002-5748-4558}$^{18}$, V.~Ashley~Villar\orcidlink{0000-0002-5814-4061}$^{35}$ \\
$^{1}$Physics and Astronomy Department, Johns Hopkins University, Baltimore, MD 21218, USA\\
$^{2}$Space Telescope Science Institute, 3700 San Martin Drive, Baltimore, MD 21218, USA\\
$^{3}$Department of Physics and Astronomy, University of Kansas, Lawrence, KS 66045, USA\\
$^{4}$Institut d'Astrophysique de Paris, CNRS-Sorbonne Universit\'e, 98 bis boulevard Arago, F-75014 Paris, France\\
$^{5}$Department of Computer Science, Johns Hopkins University, Baltimore, MD 21218, USA\\
$^{6}$School of Physics, Trinity College Dublin, The University of Dublin, Dublin, Ireland\\
$^{7}$Department of Astronomy, Oskar Klein Centre, Stockholm University, AlbaNova, SE--106~91 Stockholm, Sweden.\\
$^{8}$Steward Observatory, University of Arizona, 933 N. Cherry Ave., Tucson, AZ 85721, USA\\
$^{9}$Department of Astronomy, University of California, Berkeley, CA 94720-3411, USA\\
$^{10}$Gemini Observatory, 670 North A‘ohoku Place, Hilo, HI 96720-2700, USA\\
$^{11}$Wood Specialist in Astronomy\\
$^{12}$Department of Physics and Astronomy, Texas A\&M University, 4242 TAMU, College Station, TX 77843, USA\\
$^{13}$George P. and Cynthia Woods Mitchell Institute for Fundamental Physics \& Astronomy, College Station, TX 77843, USA\\
$^{14}$Las Cumbres Observatory, 6740 Cortona Drive, Suite 102, Goleta, CA 93117-5575, USA \\
$^{15}$Department of Physics, University of California, Santa Barbara, CA 93106-9530, USA \\
$^{16}$Department of Astronomy, University of Virginia, P.O. Box 400325, Charlottesville VA 22904-4325, USA\\
$^{17}$Dept.\ of Physics \& Astronomy, Louisiana State University, Baton Rouge, LA 70803, USA\\
$^{18}$Department of Astronomy and Astrophysics, University of California, Santa Cruz, CA 95064, USA\\
$^{19}$Department of Physics and Astronomy, Michigan State University, East Lansing, MI 48824, USA\\
$^{20}$Center for Astrophysics \textbar{} Harvard \& Smithsonian, 60 Garden Street, Cambridge, MA 02138-1516, USA \\
$^{21}$The NSF AI Institute for Artificial Intelligence and Fundamental Interactions, USA \\
$^{22}$Department of Physics and Astronomy, Rutgers the State University of New Jersey, 136 Frelinghuysen Road, Piscataway, NJ 08854, USA\\
$^{23}$3Division of Physics, Mathematics, and Astronomy, California Institute of Technology, Pasadena, CA 91125, USA\\
$^{24}$Minnesota Institute for Astrophysics, University of Minnesota, 115 Union St. SE, Minneapolis, MN 55455, USA\\
$^{25}$Department of Physics and Astronomy, Northwestern University, 2145 Sheridan Rd, Evanston, IL 60208, USA\\
$^{26}$Center for Interdisciplinary Exploration and Research in Astrophysics (CIERA), Northwestern University, 1800 Sherman Ave, Evanston, IL 60201, USA\\
$^{27}$Department of Experimental Physics, University of Szeged, H-6720 Szeged, D\'om t\'er 9., Hungary\\
$^{28}$ELKH-SZTE Stellar Astrophysics Research Group, H-6500 Baja, Szegedi {\'u}t, Kt. 766, Hungary\\
$^{29}$National Astronomical Research Institute of Thailand, 260  Moo 4, Donkaew,  Maerim, Chiang Mai, 50180, Thailand\\
$^{30}$Caltech/IPAC, Mailcode 100-22, Pasadena, CA 91125, USA\\
$^{31}$Eustace Specialist in Astronomy\\
$^{32}$Institute for Astronomy, University of Hawaii, 2680 Woodlawn Drive, Honolulu, HI 96822, USA\\
$^{33}$Department of Astronomy, University of Illinois at Urbana-Champaign, 1002 W. Green St., IL 61801, USA\\
$^{34}$DARK, Niels Bohr Institute, University of Copenhagen, Jagtvej 128, 2200 Copenhagen, Denmark\\
$^{35}$Department of Astronomy \& Astrophysics, The Pennsylvania State University, University Park, PA 16802, USA\\
}
\begin{document}

\maketitle
\clearpage
\begin{abstract}
A growing number of supernovae (SNe) are now known to exhibit evidence for significant interaction with a dense, pre-existing, circumstellar medium (CSM). SNe Ibn comprise one such class that can be characterised by both rapidly evolving light curves and persistent narrow He\one\ lines. The origin of such a dense CSM in these systems remains a pressing question, specifically concerning the progenitor system and mass-loss mechanism. In this paper, we present multi-wavelength data of the Type Ibn SN 2020nxt, including {\it HST}/STIS ultraviolet spectra. We fit the data with recently updated CMFGEN models designed to handle configurations for SNe Ibn. The UV coverage yields strong constraints on the energetics and, when combined with the CMFGEN models, offer new insight on potential progenitor systems. We find the most successful model is a $\lesssim4$~\msolar\ helium star that lost its $\sim 1\,{\rm M}_\odot$ He-rich envelope in the years preceding core collapse. We also consider viable alternatives, such as a He white dwarf merger. Ultimately, we conclude at least some SNe Ibn do not arise from single, massive ($>$30 \msolar) Wolf-Rayet-like stars.

\end{abstract}

\section{Introduction} 
\label{sec:intro}

Core-collapse supernovae (CCSNe) are the explosive endpoint of massive stars' lives. These stars typically explode in relatively low-density environments that have been self-cleared by stellar radiation pressure and low-density winds prior to explosion.  A growing number of transients, however, have now been observed to explode in a high-density, slow-moving, circumstellar medium (CSM) likely generated by progenitor pre-SN mass-loss on the order of $10^{-3}$--$10^{-1}$\,\ml \citep{smith17}.  These pre-shocked, high-density winds emit relatively narrow ($\lesssim 1000$\,km\,s$^{-1}$) lines, adding an ``n" classification to the subclass \citep[e.g., SNe~IIn, Ibn;][and those within]{smith17}.

SNe~Ibn (see \citealt{hosseinzadeh17} for a review) is one such CCSN subclass that exhibits hydrogen-poor (e.g., stripped-envelope Type Ib) spectra dominated by narrow He\one\ emission lines. Their optical light curves are distinct from those of normal SNe~Ib given their high peak luminosity and fast rise/decline. In the case of the SNe~Ibn 2006jc and 2019uo, imaging observations reveal a pre-SN outburst, potentially an extreme mass-loss event, $\sim 2$\,yr prior to explosion that may have been the origin of the CSM \citep{pastorello07,strotjohann21}. 

Recently, \citealt{fox19} proposed a potential connection between SNe Ibn and another puzzling subclass named Fast Blue Optical Transients (FBOTs), in particular AT 2018cow \citep{margutti18}. These extreme transients are defined by their light curves' relatively fast rise and decline ($< 10$ days above half-maximum), high peak bolometric luminosity ($>10^{44}$\,erg\,s$^{-1}$), and blue colour \citep[e.g.,][]{drout14}. Almost all FBOTs also show narrow lines and signs of interaction with CSM. \citet{ho21} provide a thorough review, noting that there may really be at least two distinct groups: (1) at the lower-energy end, the more common SNe~Ibn (hydrogen-poor stripped-envelope Type Ibn-like), and (2) at the high-energy end, the much less frequent FBOTs (e.g., AT2018cow; \citealt{margutti18}). There may even be other types of fast transients with signs of CSM intetaction connected to these groups, such as the H-poor {\it and} He-poor Type Icn SNe \citep{perley21,galyam22, 2022ann, Pellegrino2022}.

SNe~Ibn, SNe~Icn, and FBOTs invoke a number of questions specific to our understanding of stellar evolution and explosions. For many stripped-envelope SNe, two of the most common progentior systems proposed are a single massive Wolf-Rayet (WR) star with a high mass-loss rate \citep{gaskell86, pastorello07, smith08jc} or a close binary system \citep{podsiadlowski92}. Earlier studies of the Type Ibn SN 2006jc suggested the dense CSM was formed by a strong wind from a massive WR progenitor \citep[e.g.][]{foley07,pastorello07}. This single, massive WR star model, however, is problematic \citep{dessart22}. Most known physical mechanisms for normal stellar winds, particularly for single stars, are insufficient to produce the inferred mass-loss rates in SNe Ibn \citep[e.g., ][]{smith14}. Theory predicts that a standard WR star explosion with inner shell decelerated by CSM down to speeds of $\sim 2000$\,\kms\ should yield a superluminous SN \citep{D16_2n,dessart22}, but SNe~Ibn like SN 2006jc only have peak luminosities of $L_{\rm peak} \approx 10^{43}$\,\ergs, at least an order of magnitude fainter than SLSNe. Also, the presence of He\one\ lines at all times requires a He-dominated CSM composition, which is directly incompatible with most massive WR-star models. 

Instead, such massive CSMs require eruptive mass loss akin to luminous blue variables \citep{smith11impostor} or pre-SN binary interaction \citep{sa14} may be required. Late-time {\it Hubble Space Telescope (HST)} imaging of SN~Ibn explosion sites indicate that at least some SNe~Ibn do not originate from high-mass stars \citep{sun20, hosseinzadeh19}. 
\citet{shivvers17} use pre-explosion images to rule out a WR progenitor for the Type Ibn SN 2015G \citep{shivvers17}. Progenitor masses of $\lesssim$12~$M_\odot$ are further constrained by local stellar population studies \citep{sun22}. Furthermore, modelling of Type Ibn spectra \citep{dessart22} showed that Type~Ibn SN progenitors have final masses of only $\lesssim$5~$M_\odot$, much lower than those for WR stars.

Recently, the theoretical community has proposed a wide range of new physical scenarios. Binary models are successful at explaining at least some of the observed CSM characteristics of SNe~Ibn by invoking nuclear flashes in low-mass He stars \citep{woosley_he_19, dessart22} or repeated mass-transfer episodes in short-period binaries \citep{dewi_pols_03,pols_94,tauris_ulstr_13, langer_araa_mdot_12}. Other proposed explosion models include a double white dwarf merger \citep{Lyutikov&Toonen2019, dessart22, metzger22}, a Thorne-Zytkow-like object (TZlO) as a result of head-on collision between white dwarf and neutron star (NS) collapse to a NS or a black hole as a central engine \citep[see][]{Pascalidis+11b}, a millisecond-period magnetar powering an ultrarelativistic gamma-ray-burst jet (e.g., \citealt{metzger11}), and other potential extreme scenarios, such as intermediate mass black holes \citep[e.g.,][]{perley21}. Whether or not multiple mechanisms are at work, more observations are needed to make any connection between this variety of mechanisms and progenitors and the diversity of SNe~Ibn. Given the large number of possible interaction configurations, one may even expect a continuum of events covering the parameter space from SNe~Ibn/Icn to FBOTs \citep[e.g.,][]{dessart22}. Stellar evolution, in particular if one invokes binarity, can produce a large set of evolutionary paths.

Unlike many of the other SN subclasses, SNe~Ibn do not yet have direct observational evidence of their quiescent progenitor systems. Instead, the community has had to rely on indirect evidence, mostly consisting of optical spectroscopy/photometry and host-galaxy studies \citep[][and those within]{hosseinzadeh19}. These observations have highlighted the need for multiwavelength, multiepoch observations, especially in the ultraviolet (UV) \citep[e.g.,][]{dessart22}. The bolometric light curve, built from UV, optical, and near-infrared (NIR) observations, is essential to constrain the energy budget of these transients and identify a multicomponent emission source. 
In addition, the degeneracy in the light-curve properties requires the additional wealth of information encoded in spectra (e.g., composition, line-profile properties such as morphology, width, or strength).

In this paper we present multiwavelength observations of SN~Ibn~2020nxt, including UV, optical, and NIR. Section \ref{sec:obs} presents the observations, while \S \ref{sec:analysis} discusses the modelling of the spectral series.  In \S \ref{sec:discussion} we examine line diagnostics and their implications for the progenitor system and evolution channel.  
Finally, \S \ref{section:conclusion} provides a conclusion of the work and its implications for the future. Throughout the paper, we adopt a flat $\Lambda$CDM cosmological model with H$_0 = 73$ km s$^{-1}$ Mpc$^{-1}$ and $\Omega_m = 0.27$ \citep[][]{riess22}.

\section{Observations}
\label{sec:obs}

\begin{figure*}
\begin{center}
\includegraphics[width = 0.95\textwidth]{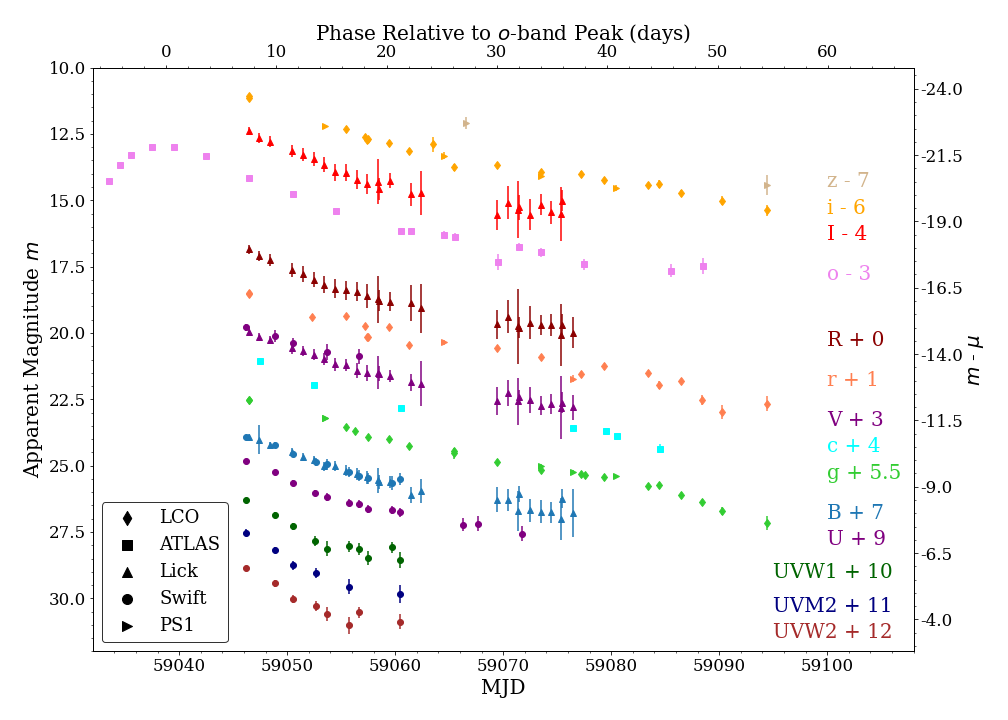}
\caption{Multiband light curves of SN 2020nxt within the first $\sim70$\,days. The rest-frame phases labelled at top of the plot are relative to the inferred time of peak brightness in the ATLAS $o$ band on MJD 59038.76. For clarity each band has been shifted as labeled in the annotations. 
}
\end{center}
\label{fig:lc}
\end{figure*}

SN 2020nxt was first discovered by the Asteroid Terrestrial-impact Last Alert System (ATLAS; \citealp{tonry2018atlas}) on 2020 July 3 12:53:16.8 UTC (MJD 59033.537) in the $o$ band with $m_o = 17.192$\,mag \citep[][]{discovery}, and was spectroscopically classified as an SN~Ibn on July 14 with the Liverpool Telescoope \citep[][]{classification}. The last nondetection on SN 2020nxt was on 2020 July 1 14:29:46 UTC (MJD 59031.604) with a limiting magnitude $m_o\lesssim19.67$.  We adopt a redshift of $z=0.0218\pm 0.003$ measured from the centroid of He~I $\lambda$7065 through all phases. This corresponds to a distance of $91.1\pm1.2$\,Mpc and a distance modulus of $\mu = 34.798\pm0.032$\,mag. The Milky Way extinction is relatively low toward this direction, with $E(B-V)_{\rm MW} = 0.067$\,mag \citep{2011ApJ...737..103S}.
The peak of \atlas\ $o$-band is on MJD 59039.485 with $m_{o{\rm , peak}} = 15.985$\,mag. We applied a quadratic fit to the \atlas\ $o$-band light curve from MJD 59035 to 59045 and estimated the $o$-band peak to be on 2020 July 8 18:14:24 UT (MJD 59038.76). 



\subsection{UV, Optical, Near-IR Imaging \label{sec:UVOIRimaging}}

The \atlas\ images are processed as described by \cite{tonry2018atlas}, and then photometrically and astrometrically calibrated using the RefCat2 catalogue \citep{tonry18ref}. Template generation, image-subtraction procedures, and photometric measurements are carried out following \cite{smith2020}. 
The \atlas\ light curves are then cleaned, averaged and converted into AB magnitude system using a suite of routines described by \cite{RestSofia21}\footnote{\url{https://github.com/srest2021/atlaslc}}.
The first cut uses the $\chi^2$ and uncertainty values of the point-spread-function (PSF) fitting to clean out bad data. We then obtain forced photometry of 8 control light curves located in a circular pattern around the SN with a radius of $17''$. The flux of these control light curves is expected to be consistent with zero within the uncertainties and any deviation from that would indicate that there are either unaccounted systematics or underestimated uncertainties. We search for such deviations by calculating the 3$\sigma$ cut weighted mean of the set of control light-curve measurements for a given epoch (for a more detailed discussion, see Rest et al., in prep.). The weighted mean of these photometric measurements is expected to be consistent with zero and, if not, we flag and remove those epochs from the SN light curve. This method allows us to identify potentially bad measurements in the SN light curve without using the SN light curve itself. We then bin the SN 2020nxt light curve by calculating a 3$\sigma$ cut weighted mean for each night (typically, \atlas\ has 4 epochs per night), excluding the flagged measurements from the previous step. We find that this method successfully removes outliers from the SN light curve.

Follow-up observations of SN~2020nxt were also performed by the 0.76\,m Katzman Automatic Imaging Telescope (KAIT) as part of the Lick Observatory Supernova Search (LOSS; \cite{Filippenko2001}), as well as the 1\,m Nickel telescope at Lick Observatory. $B$, $V$, $R$, and $I$ multiband images of SN~2020nxt were obtained with both telescopes, and additional $Clear$-band images (close to the $R$ band; see \cite{Li2003}) were obtained with KAIT. 
All images were reduced using a custom pipeline\footnote{https://github.com/benstahl92/LOSSPhotPypeline} detailed by \cite{Stahl2019}. The image-subtraction procedure was applied in order to remove the host-galaxy light, using additional images obtained after the SN had faded below our detection limit. PSF photometry was obtained using DAOPHOT \citep[][]{stetson87} from the IDL Astronomy User’s Library\footnote{http://idlastro.gsfc.nasa.gov/}. Several nearby stars were chosen from the Pan-STARRS1\footnote{http://archive.stsci.edu/panstarrs/search.php} catalogue for calibration; their magnitudes were first transformed into the Landolt \citep{landolt92} system using the empirical prescription presented by Eq.~6 of \citet{Tonry2012}, and then transformed to the KAIT/Nickel natural system. Apparent magnitudes were all measured in the KAIT4/Nickel2 natural system. The final results were transformed to the standard system using local calibrators and colour terms for KAIT4 and Nickel2 \citep{Stahl2019}.

\defcitealias{2017ApJS..233...25A}{SDSS Collaboration 2017}
SN~2020nxt was observed with Las Cumbres Observatory \citep{2013PASP..125.1031B} as part of the Global Supernova Project, using the Sinistro camera on the 1\,m telescope at McDonald Observatory (TX, USA) and the Spectral camera on the 2\,m Faulkes Telescope North (FTN) at Haleakal\=a Observatory (HI, USA). These images were preprocessed using BANZAI \citep{2018SPIE10707E..0KM}. We subtracted reference images taken with the same 1\,m telescope on 2022 Sep. 17 UTC using PyZOGY \citep{2017zndo...1043973G} and measured PSF photometry on the difference images using \texttt{lcogtsnpipe} \citep{2016MNRAS.459.3939V}. \textit{UBV} Vega magnitudes are calibrated to \cite{1983AJ.....88..439L} standard fields taken on the same nights with the same telescopes, and \textit{gri} AB \citep{oke83} magnitudes are calibrated to the Sloan Digital Sky Survey (SDSS; \citetalias{2017ApJS..233...25A}).

We also observed SN\,2023bee with the PanSTARRS1 (PS1) telescope \citep{2016arXiv161205560C} in $griz$ bands as part of the Young Supernova Experiment \citep{Jones2021,Aleo2022}. The PS1 images are reduced by the PS1 image Image Processing Pipeline (IPP) \citep{2020ApJS..251....3M,2020ApJS..251....5M,2020ApJS..251....6M,2020ApJS..251....4W}, and then calibrated using the Pan-STARRS DR1 catalog \citep{2020ApJS..251....7F}.

UV and additional optical imaging was performed with the {\it Neil Gehrels Swift Observatory} \citep{Gehrels_2004, Roming_2005}. Observations began on 2020 July 16 at 02:38:13 UTC. Reductions were performed using the methods outlined by \citet{Brown_2009} and the pipeline of the {\it Swift} Optical Ultraviolet Supernova Archive (SOUSA; \citealp{Brown_2014}). The count-rate flux from the host galaxy within the source aperture was subtracted using an observation on 2021 Jan. 22.

\begin{figure}
\begin{center}
\includegraphics[width=0.5\textwidth]{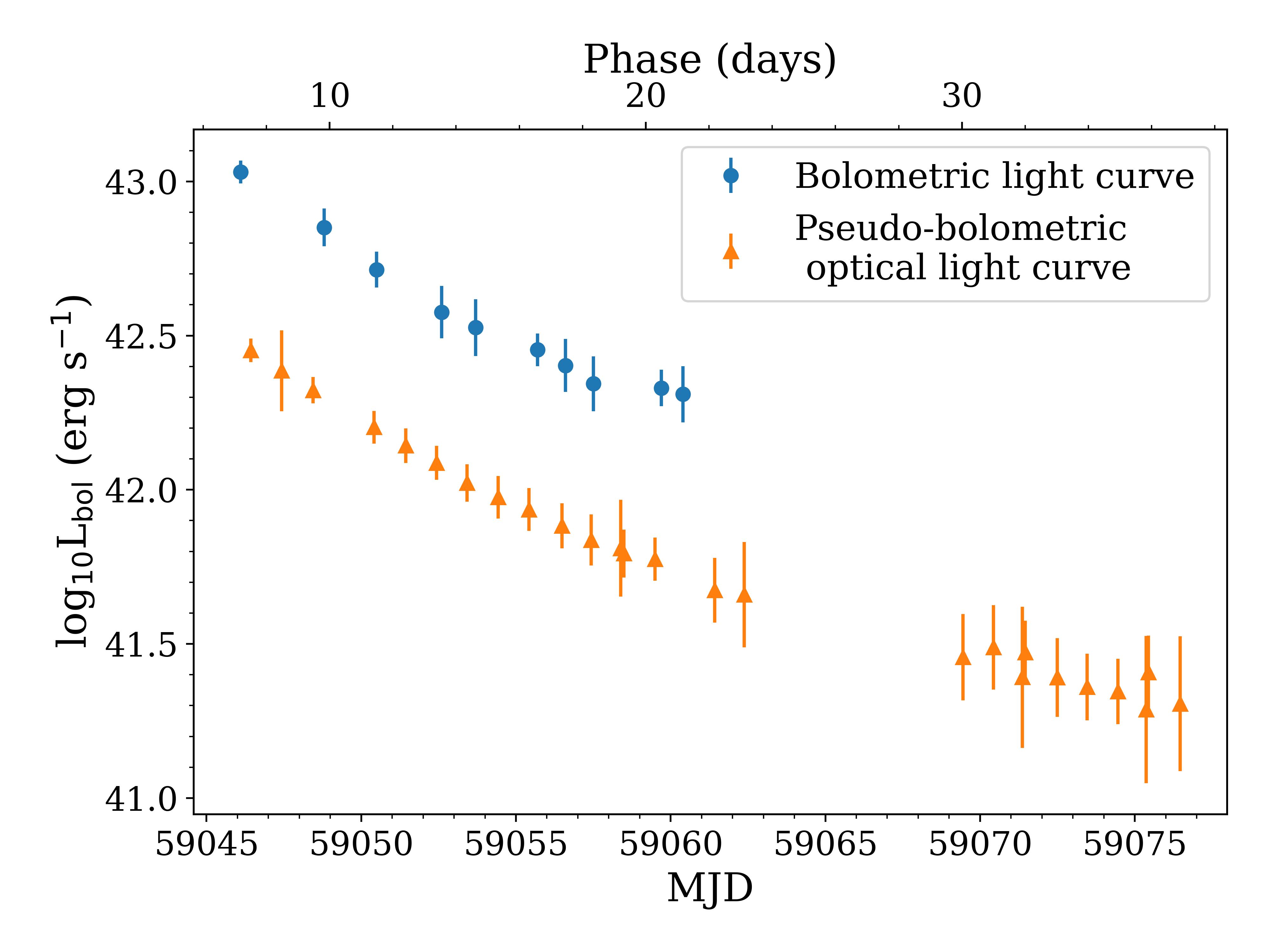}
\hspace{-3in}
\end{center}
\caption{Bolometric and pseudobolometric light curves of SN~2020nxt. The bolometric light curve (blue) is estimated by fitting a blackbody SED to the UV--optical light curve when UV data from {\it Swift} are present. The pseudobolometric light curve (orange) is estimated by integrating through the $BVRI$ bands as an indicator of the bolometric light-curve evolution at later phases. The phases are relative to $o$-band maximum brightness. }
\label{fig:bololc}
\end{figure}

The multiband light curves are shown in Figure~\ref{fig:lc}. 
We further use the \textsc{Superbol} pipeline\citep{superbol} to calculate the bolometric light curve from our optical and UV light curves, as shown in Figure~\ref{fig:bololc}. In order to align measurements in different filters on the same epochs, we use the $B$ band as the reference filter and interpolate and extrapolate the light curves in other bands with second-order polynomials  between MJD 59045 and 59065. We then estimate the full bolometric light curve by fitting a blackbody SED. For the later phases when no UV data are present, we also calculated the pseudobolometric light curve by integrating the photometry in $BVRI$ bands as a reference to the bolometric luminosity evolution, as shown in Figure~\ref{fig:bololc}. 


\subsection{Host Galaxy}
\begin{figure}
    \centering
    \includegraphics[width=\columnwidth]{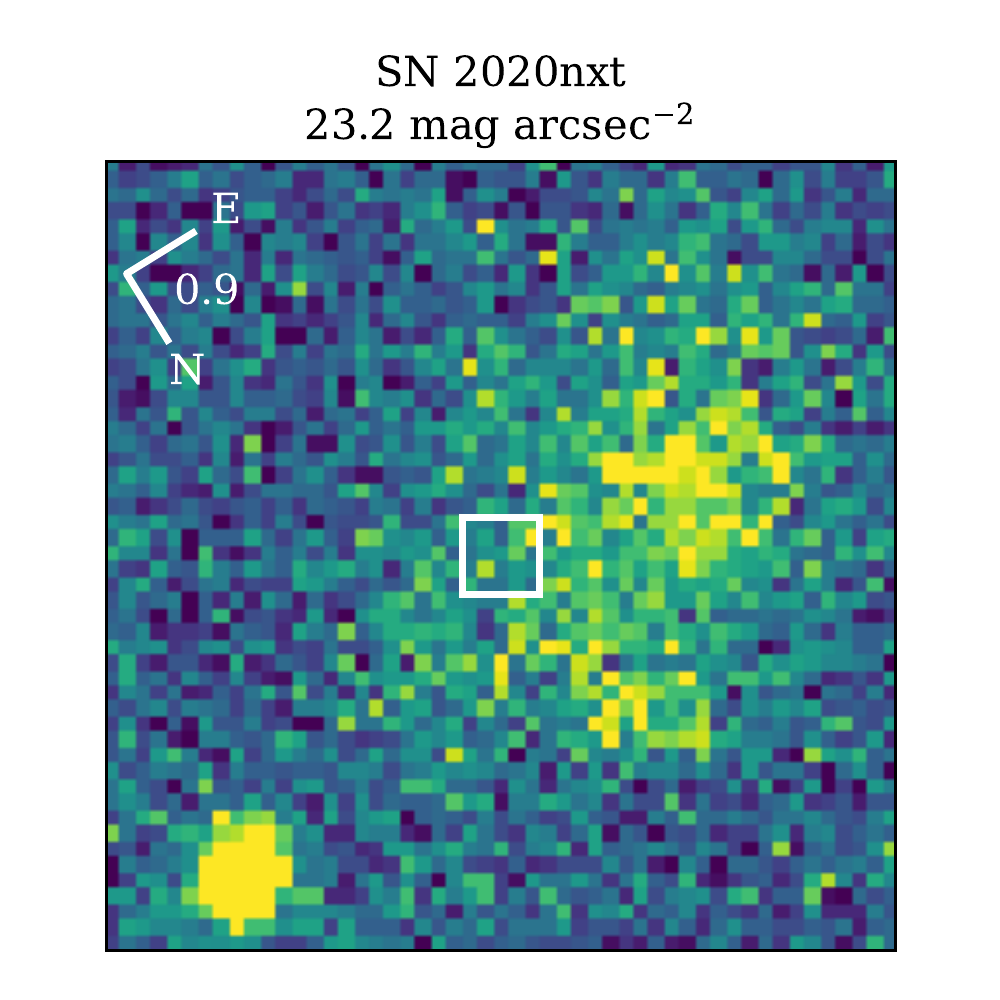}
    \caption{A $u'$-band image of the host galaxy of SN~2020nxt. By measuring the UV flux within a $5 \times 5$ pixel$^2$ ($0.9 \times 0.9$\,kpc$^2$) aperture centred on the SN location (white square), we derived a surface brightness of 23.2\,mag\,arcsec$^{-2}$ and an SFR density of $0.019\, {\rm M}_\odot\mathrm{\ yr^{-1}\, kpc^{-2}}$, near the median for SNe~Ibn \citep{hosseinzadeh19}.}
    \label{fig:host}
\end{figure}

\defcitealias{2017ApJS..233...25A}{SDSS Collaboration (2017)}
\noindent SN~2020nxt exploded $4.5''$ (2.0\,kpc projected) west of the spiral galaxy WISEA~J223736.70+350006.5. Following the methodology of \cite{hosseinzadeh19}, we downloaded a $u'$-band image of this galaxy from the \citetalias{2017ApJS..233...25A} and measured the surface brightness within a $5$\,pixel $\times 5$\,pixel ($0.9$\,kpc $\times 0.9$\,kpc) aperture centred on the SN location (Fig.~\ref{fig:host}). The result, $\sigma_{u'} = 23.2 \pm 0.2\, \mathrm{\ mag\, arcsec^{-2}}$, implies a star-formation-rate (SFR) density of $\Sigma_\mathrm{SFR} = 0.019\, {\rm M}_\odot\mathrm{\ yr^{-1}\, kpc^{-2}}$, according to the relationship of \cite{1998ARA&A..36..189K}. This places the site of SN~2020nxt near the median SFR density among SN~Ibn \citep{hosseinzadeh19} and other CCSN \citep{2018ApJ...855..107G} hosts.

\subsection{Pre-explosion Imaging}

\begin{figure}
\begin{center}
\hspace{-0.2in}
\includegraphics[width=0.5\textwidth]{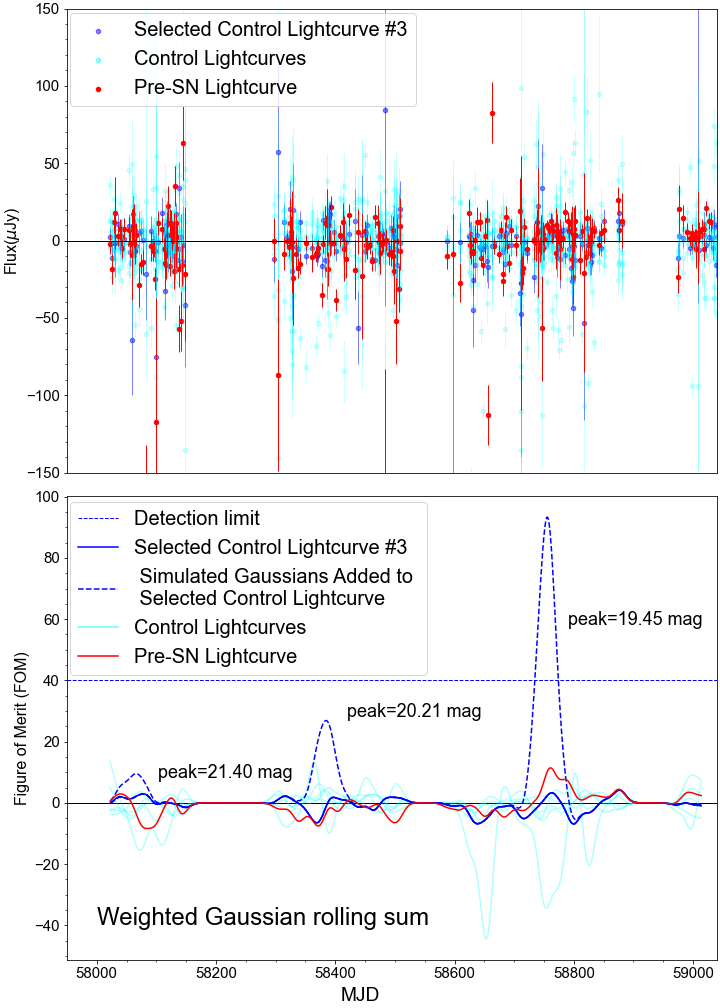}
\end{center}
\caption{\textbf{Top:} The pre-SN light curve (red) in the $o$ band compared with the control light curves (blue) taken from the background around the position of SN~2020nxt. One of the control light curves has been highlighted (cyan) for clarity. \textbf{Bottom:} The weighted sum of the S/N convolved with a rolling Gaussian with standard deviation $\tau_G = 10$\,days. In addition, we add artificial Gaussians with different configurations to the control light curves to determine the detectability of different types of eruptions. Dashed line in cyan is an example showing three Gaussians with $\tau^\prime_G = 15$\,days and different magnitudes added onto the selected control light curve. The horizontal dashed line denotes the adopted limit of 15 of the Figure of Merit as determined by its maximum in the control light curves.}
\label{fig:preeruption_rollinggaussian}
\end{figure}

Given the pre-SN outburst observed two years prior to SN~2006jc \citep{foley07,pastorello07}, we searched the pre-explosion data of SN~2020nxt for a similar brightening event. \atlas\ monitored the field of SN~2020nxt in the $c$ and $o$ bands for $\sim 3$\,yr prior to explosion. Figure \ref{fig:preeruption_rollinggaussian} shows pre-explosion \atlas\ light curves generated by forced photometry at the position of the SN.

In order to determine whether there are pre-explosion outbursts within a light curve, we aim to account for contamination caused by factors such as instrument and reduction artifacts or nearby bright objects, while simultaneously remaining sensitive to faint eruptions.

We first define a figure of merit (FOM) as the flux-to-uncertainty ratio convolved with the rolling Gaussian in order to emphasise the signature of an eruption. As eruptions typically have timescales of $\tau_{\rm eruption}$ between 5 and 50\,days, we expect that a real eruption should exhibit consecutive significant detections above zero, even if its individual detections may not exceed 3$\sigma$ above zero. We convolve the signal-to-noise ratio (S/N) with a rolling Gaussian characterised by a standard deviation $\tau_{G}$ that is similar to the timescale of the target eruption $\tau_{\rm eruption}$. Such a convolution strongly emphasises the signal of any real (though faint) eruption. 

\begin{figure}
\begin{center}
\hspace{-0.3in}
\includegraphics[width=0.5\textwidth]{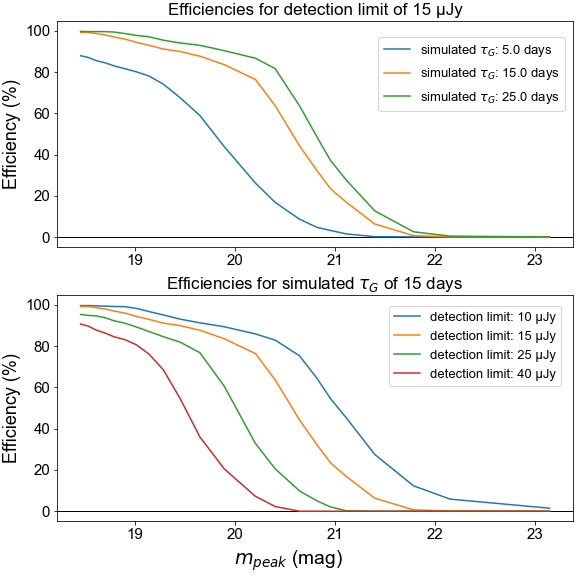}
\end{center}
\caption{The detection efficiencies of eruptions with different amplitudes (\textit{top}) and $\tau_G$ under different detection limits.}
\label{fig:preeruption_efficiency}
\end{figure}

We next guard against possible contamination and false detections within the pre-SN light curve. As described in Section~\ref{sec:UVOIRimaging}, we utilise a series of control light curves to establish a detection limit for real pre-explosion outbursts. Since we can assume that each control light curve will not contain any pre-explosion outbursts, we apply the aforementioned rolling Gaussian to the control light-curve flux so as to determine the control FOM, above which any real explosion are expected to rise. We can use this to set a detection threshold FOM$_{\rm limit}$: If none of the FOMs of the control light curves are above FOM$_{\rm limit}$, then we can expect to have very little contamination. 

The top panel of Figure~\ref{fig:preeruption_rollinggaussian} displays the pre-explosion SN~2020nxt light curve in red, as well as eight control light curves in blue. A single control light curve has been selected for illustration and coloured in cyan. The lower panel shows the FOM for each light curve in the corresponding colour, as well as the derived detection threshold of FOM$_{\rm limit}=15$ for a possible eruption within SN~2020nxt. In order to confirm that eruptions of certain peak magnitudes would be detected above the detection limit, we have added three simulated Gaussian bumps with increasing peak magnitudes to the selected cyan control light curve. As depicted, the two largest peaks with magnitudes of 20.21 and 19.45 rise above the detection limit. We can thus assume that eruptions of these magnitudes or brighter would be underscored in the SN light curve after convolution by the rolling Gaussian.

Finally, we calculate the efficiency of our eruption-detection method. We begin by applying a series of simulated Gaussian eruptions to all control light curves, randomly drawing the width of the Gaussian from a possible range of 3 to 50\,days, as well as randomising the location of the peak within seasonal observation windows. We determine the efficiency by applying the same detection method to each simulated eruption, and then scanning the light curve for FOMs above a predetermined detection limit and within 1$\sigma$ of the location of the added Gaussian's peak. Detections which fulfill this criterion are marked as successful and compared to the total number of added simulated eruptions. 

The upper panel of Fig.~\ref{fig:preeruption_efficiency} displays efficiencies for different input Gaussian width ranges using a detection limit of 15. As expected given the cadence of ATLAS, the efficiency drops significantly for input Gaussian widths $\tau_G$ below 25\,days. The lower panel additionally examines the efficiencies for a range of possible detection thresholds, calculated from input Gaussians with widths $\tau_G = 15$\,days. Using a detection limit of 15\,$\mu$Jy, approximately 80\% of eruptions with peak magnitudes of 20 are successfully recovered.  

The maximum of FOM of the pre-explosion light curve of SN~2020nxt is 11.4 on MJD 58760.5, $\sim270$\,days prior to the first detection in the $o$ band. Given that this value is smaller than FOM$_{\rm limit} = 15$, we can safely conclude that no pre-explosion eruption has been found with a detection limit of $m\lesssim20$\,mag, or $M\lesssim-14.8$\,mag. This upper limit is marginally smaller than the pre-explosion eruption detected for SN~2006jc which has $M=-14$ \citep{Nakano2006}. 

\begin{table*}
\caption{Log of spectroscopic observations of SN 2020nxt. The phases are relative to $o$-band maximum brightness. 
}
\centering
\begin{tabular}{lcccccc}
\hline
\hline
MJD & Phase [days] & Telescope & Instrument & Grism/Grating & Wavelength Range [{\AA}]\\
\hline
59044.1 & +5.3 & LT & SPRAT &  ---  & 4020-7580 \\
59046.5 & +7.6 & FTN & FLOYDS &  ---  & 3500-10000\\
59048.4 & +9.4 & Shane & Kast &  ---  & 3618-10712\\
59049.0 & +10.0 & NOT & ALFOSC &  --- & 4003-9689\\
59049.5 & +10.5 & Shane & Kast &  --- & 3306-10496\\%
59049.7 & +10.7 & HST & STIS &  G140L+G230L& 1138-3180  \\
59052.5 & +13.4 & FTN & FLOYDS &  ---  & 3500-10000\\
59053.4 & +14.3 & Keck I & LRIS &  ---  & 3302-10097\\%
59055.0 & +15.9 & NOT & ALFOSC &  --- & 4000-9635\\
59058.4 & +19.2 & Shane & Kast &  ---  & 3616-10688\\%
59058.7 & +19.5 & HST & STIS &  G230L& 1570-3180  \\
59061.5 & +22.3 & FTN & FLOYDS &  ---  & 3500-10000\\
59062.1 & +22.8 & NOT & ALFOSC & --- & 4002-9624\\
59065.5 & +26.2 & FTN & FLOYDS & ---  & 3500-10000\\
59070.6 & +31.2 & Keck I & LRIS & ---  & 3505-10493\\%
59072.4 & +32.9  & Shane & Kast & ---  & 3618-10714\\%
59074.5 & +35.7 & Keck 2 & NIRES & --- & 9650-24670\\ 
59083.6 & +43.9 & Keck I & LRIS & 600/4000+400/8500 & 3156-10280\\ 
59084.6 & +44.9 & Keck I & LRIS & 600/4000+400/8500 & 3202-10147\\ %
\hline
\end{tabular}
\label{tab:spectra}
\end{table*}

\begin{figure*}
\begin{center}
\hspace{-0.4in}
\includegraphics[scale=0.5]{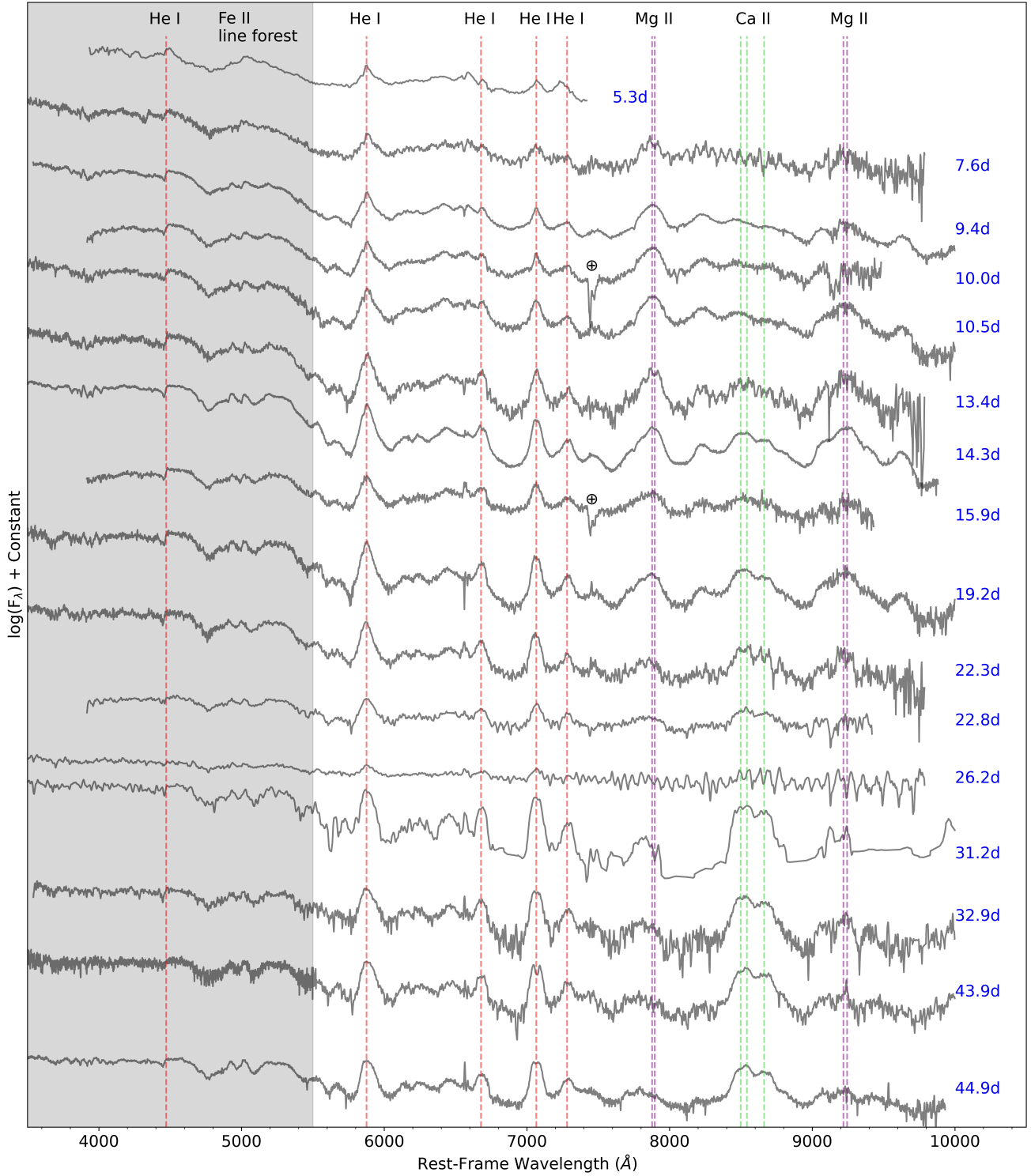}
\caption{Optical spectral series of SN~2020nxt in the rest frame shown in log scale. Vertical dashed lines indicate the wavelengths of some of the strongest lines. The grey shadowed region below $5500$\AA\ is dominated by the Fe\two\ line forest and hard to identify any line features except for He\one\ 4471\,\AA.
All of the spectra have been normalised to the continuum in the range 6100--6300\,\AA. Phases relative to the $o$-band peak at MJD 59038.76 are labelled to the right side of each spectrum.  
}
\label{fig:Opticalspec}
\end{center}
\end{figure*}

\subsection{Optical and Near-IR Spectroscopy}

Optical spectroscopy was obtained with several telescopes, summarised in Table \ref{tab:spectra} and plotted in Figure~\ref{fig:Opticalspec}. FLOYDS spectra from Faulkes Telescope North (FTN) were observed with a $2''$-wide slit and reduced using the \texttt{floyds\_pipeline}\footnote{\url{https://github.com/LCOGT/floyds_pipeline}} \citep{2014MNRAS.438L.101V}. Observations with the Nordic Optical Telescope \citep[NOT; ][]{djupvik2010} 
on La Palma used the Alhambra Faint Object Spectrograph (ALFOSC)
\footnote{Programs 61.604 \& 61-501, P.I. J. Sollerman}.  
All spectra were obtained using the parallactic angle \citep{Filippenko1982} and with airmass $< 1.3$. A slit width of 1\farcs0 and Grism 4 were used for all observations.
Three of the spectra have excellent S/N, while the fourth was affected by calima dust. The NOT spectra were reduced using the \texttt{Pypeit} pipeline
\citep{Prochaska2020,Prochaska2020b}. 
The classification spectrum was obtained using the SPectrograph for the Rapid Acquisition of Transients (SPRAT; Piascik et al. 2014) on the 2\,m Liverpool Telescope (Steele et al. 2004). SPRAT uses a $1.8''$ slit and covers a wavelength range of $\sim 4000$--8000\,\AA.

Additional optical spectra were obtained through the Young Supernova Experiment (YSE) \citep{jones21,Coulter2022,Coulter2023} with the Keck-I 10\,m telescope using LRIS \citep{Oke1995}, and also with the Lick 3\,m (Shane) telescope using the Kast spectrograph \citep{miller1993lick}. They were reduced through the {\tt UCSC Spectral Reduction Pipeline}\footnote{\url{https://github.com/msiebert1/UCSC\_spectral\_pipeline}} \citep{Siebert20}, a custom data-reduction pipeline based on procedures outlined by \citet{Foley03}, \citet{Silverman2012}, and references therein.  The two-dimensional (2D) spectra were bias-corrected, flat-field corrected, adjusted for varying gains across different chips and amplifiers, and trimmed.  1D spectra were extracted using the optimal algorithm \citep{Horne86}.  The spectra were wavelength-calibrated using internal comparison-lamp spectra with linear shifts applied by cross-correlating the observed night-sky lines in each spectrum to a master night-sky spectrum.  Flux calibration and telluric correction were performed using standard stars at a similar airmass to that of the science exposures.  We combine the sides by scaling one spectrum to match the flux of the other in the overlap region and use their error spectra to correctly weight the spectra when combining.  More details of this process are discussed elsewhere \citep{Foley03, Silverman2012, Siebert20, 2022ann}.

A near-infrared (NIR) spectrum of SN~2020nxt was obtained on 2020 Aug. 13 (MJD 59074.5) with the Near-Infrared Echellette Spectrometer (NIRES) \citep{Wilson_2004} mounted on the Keck-II 10\,m telescope, with slit width of $0.55''$. 
The SN was observed in an ABBA dithering pattern for sky subtraction, and reduced utilising the \texttt{Spextool} software package \citep{Cushing_2004}. The telluric absorption corrections were done with the \texttt{xtellcor} software. The log of NIR spectroscopic observation is given in Table \ref{tab:spectra}, and the spectrum is presented in Figure~\ref{fig:NIR}.

All spectral data and corresponding information will be made available via WISeREP\footnote{\href{https://wiserep.weizmann.ac.il}{https://wiserep.weizmann.ac.il}} \citep{YaronWiserep}.


\subsection{HST/STIS UV Spectroscopy}\label{section:hst}

\begin{figure*}
\begin{center}
\hspace{-0.4in}
\includegraphics[width=1.0\textwidth]{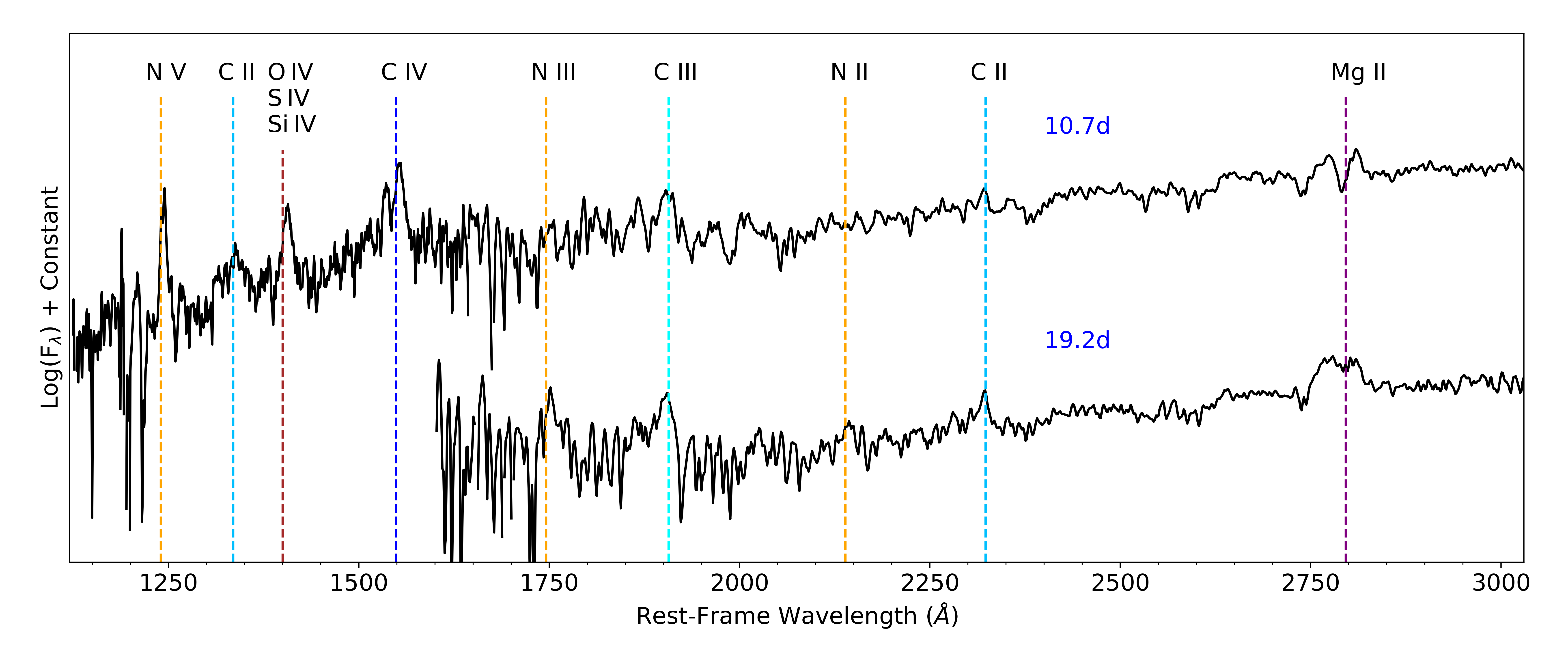}
\caption{UV spectra of SN~2020nxt taken with {\it HST} STIS at phase +10.0 and +14.3\,days past maximum. The spectra have been smoothed with a Gaussian kernel and plotted on a log scale for clarity, and the prominent line features are identified with vertical lines.}
\label{fig:UVspec}
\end{center}
\end{figure*}

SN 2020nxt was observed twice with the {\it HST}/STIS as part of program GO-15834 (PI O. Fox), as summarized in Table \ref{tab:spectra} and plotted in Figure \ref{fig:UVspec}.  The 1D spectrum for each observation is extracted using the CALSTIS custom extraction software stistools.x1d.  The default extraction parameters for STIS are defined for an isolated point source. For both G140L and G230L the default extraction box width is 7 pixels and the background extraction box width is 5 pixels. The UV and optical spectra are then flux-calibrated to the {\it Swift} UVOT UVM2 and $B$-band photometry.

\subsection{Line Analysis}  

\begin{figure*} 
\centering
\includegraphics[width=1.0\textwidth]{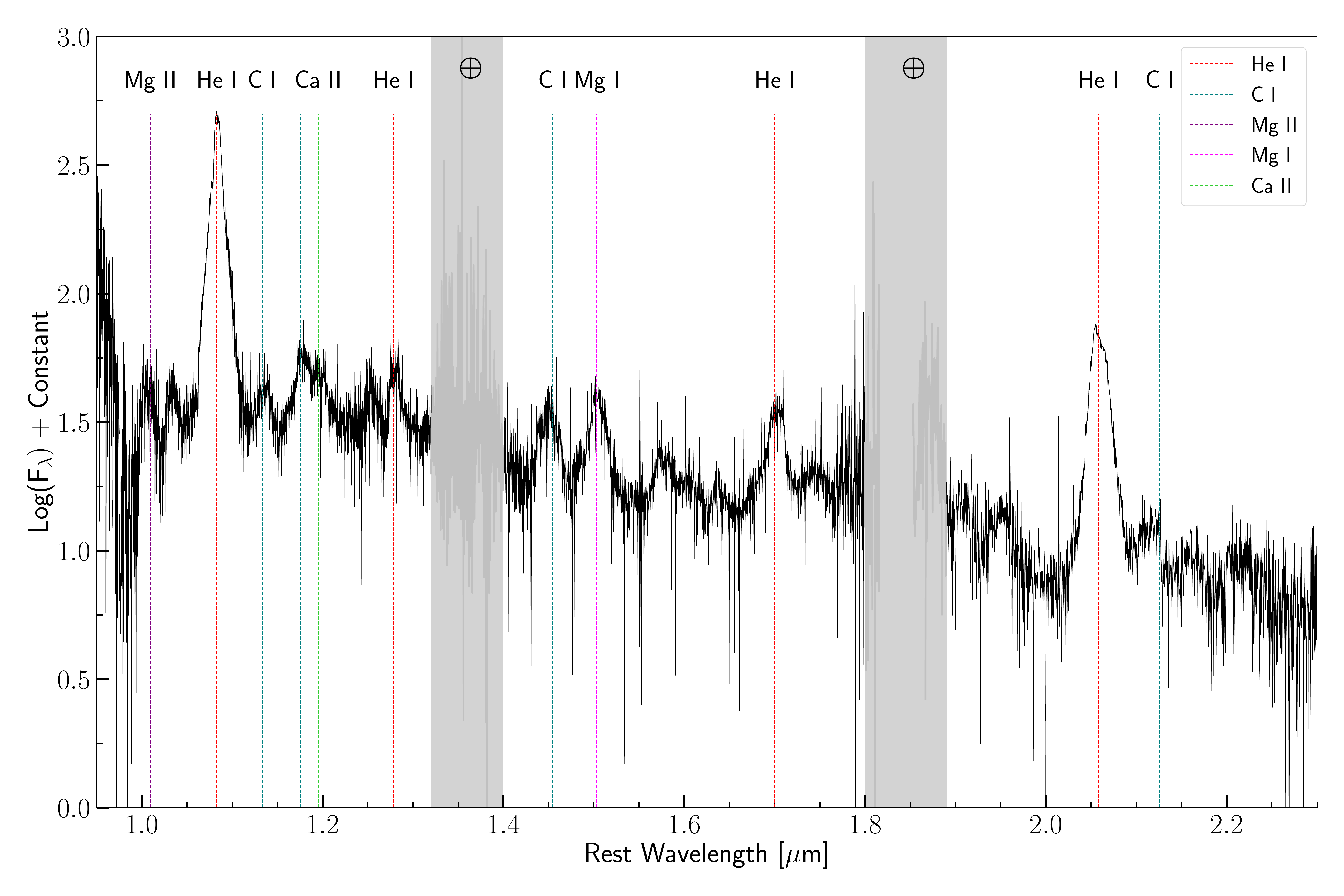}
    \caption{The NIR spectrum of SN~2020nxt obtained on 2020 Aug. 13 at phase +35.7\,days past maximum brightness with the Keck-II telescope. The two grey bands mark regions that have high telluric absorption. The flux is plotted on a log scale, and the prominent features in the spectrum are identified.}
    \label{fig:NIR}
\end{figure*}

\indent Most of the identified lines have been marked in Figs.~\ref{fig:Opticalspec}, \ref{fig:NIR}, and~\ref{fig:UVspec}. In this section we discuss the evolution of the most prominent lines.
The He features are strong throughout all phases, especially He\one\ 5876, 7065\,\AA. Mg\two\ doublets around 7880\,\AA\ and 9230\,\AA\ are prominent within the first 20\,days but gradually weakened and are hardly visible afterward. On the other hand, the Ca\two\ triplet around 8600\,\AA\ is not visible at the earliest phases, but appeared $\sim 14$\,days after the first detection and gradually became prominent at later phases. Figure~\ref{fig:lineVel} shows the selected line profiles of He\one\ 5876, 7065\,\AA, Mg\two\ doublets, and Ca\two\ triplets throughout all phases in velocity space. Despite the change in the line flux, the variation in the centroid velocities of Ca\two\ and He\one\ lines are negligible. The FWHM of these lines is $\lesssim 2000$\,km\,s$^{-1}$ throughout all phases. 

In the far-UV, a number of highly ionised emission features are present, including N\five\ 1240\,\AA, C\two\ 1335\,\AA, and the C\four\ resonance doublet at 1548, 1550\,\AA. There is another strong emission feature at 1400\,\AA\ and could originate from O\four, S\four, or Si\four. In the near-UV, the only strong features during the early phase are the Mg\two\, doublets around 2800\,\AA, while C\three\ 1907\,\AA\ and C\two\ 2323\,\AA\ can barely be seen. In the NIR spectrum taken +35.7\,days post-peak, the strongest features are the broad emission lines of He\one\ 1.083\,$\mu$m and 2.0581\,$\mu$m, while other relatively weak metal emission lines including C\one, Ca\two, and Mg\one\ can also be identified.  




\section{Analysis}
\label{sec:analysis}


\subsection {The Type Ibn Supernovae Model}
   
In order to constrain the mass of the progenitor star and ejecta, we make a comparison between the state-of-the-art simulation with \cmfgen\ and our spectral series of SN~2020nxt. The radiative-transfer simulations presented here include some of the original calculations presented by \citet[]{dessart22} (hereafter D22) as well as new calculations with different parameters. For the SN~Ibn scenario, the context is that of low- or moderate-energy, low-mass ejecta interacting with $\sim 1$\,\msun\ He-rich CSM. The progenitor for this scenario could be a low-mass binary He-rich star losing its He-rich envelope just before core collapse, followed by the evolution of the remaining star until core collapse and a weak neutrino-driven explosion --- such events are predicted from stellar-evolution models in this mass range \citep{woosley_he_19}. This envelope loss could arise, for example, through a Case BB unstable mass-transfer episode \citep{pols_94,dewi_pols_03} or through a nuclear flash as may arise in single or binary stars having a small core mass \citep{WH15,woosley_he_19}. Our numerical approach applies best after bolometric maximum light, when the inner and outer shells have been essentially entirely swept up (i.e., the interaction is essentially over) into a dense narrow shell (i.e., $\delta V / V \ll 1$, where $V$ is the ejecta velocity) that evolves in a quasisteady state and expands ballistically while the power arises from the residual interaction. 
In this work we treat directly the interaction shock power in the radiative transfer with \cmfgen\ to better accounting for greater non local thermodynamic equilibrium (NLTE) effects, H-deficient interactions and stronger role played by lines, under which the methodology in the study of SN~1994W do not work well \citep{D15_2n, D16_2n}. With this approach, we lose a little on the the physical consistency of the treatment on hydrodynamics, but we gain significantly on the treatment of the gas and the interaction between radiation and matter. 

A further assumption of D22 is to treat the shock power released in the complicated, inherently 3D interaction region in a similar way to the $\gamma$-ray energy deposition from decay power. In practice, the model injects the power within the dense shell and ignores any distinction between the relative contributions from the reverse and forward shocks. That is, it focuses on that part of the total shock power that is thermalised within the dense shell (this might be 1\%, 10\%, or 100\% of the total power produced by the shock) and injects that power in the form of high-energy electrons. Then, using the nonthermal solver in \cmfgen, the model computes the degradation of these high-energy electrons as they collide with electrons, ions, and atoms in the plasma. From this degradation spectrum, the influence of nonthermal electrons on the temperature, ionisation, and excitation is determined. The overall approach in \cmfgen\ is therefore analogous to the treatment of radioactive decay in models of standard SNe. 

\begin{figure*}
\begin{center}
\includegraphics[width=\hsize]{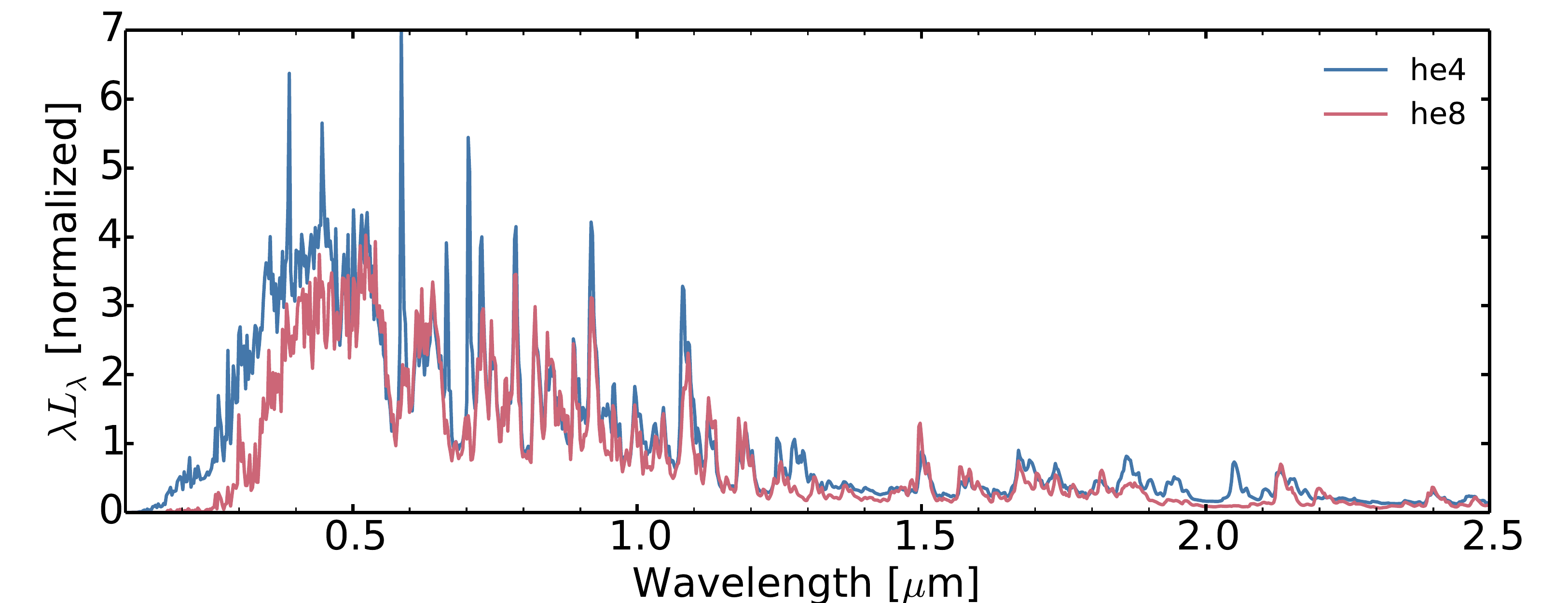}
\caption[]{Two different model spectra illustrating the impact of the He content and ejecta mass on the SED for a dense shell moving at 2000\,\kms, located at a radius of $3 \times 10^{15}$\,cm and with an injected power of $2 \times 10^{42}$\,erg\,s$^{-1}$. The models shown are he4 (total mass of 1.62\,\msun, with 0.92\,\msun\ of He) and he8 (total mass of 3.95\,\msun, with 0.84\,\msun\ of He) from D22. The model with higher mass and lower He content shows weak or no He\one\ lines (e.g., at 3888, 4471, 5875, 6678, 7065, 10,830, or 20,581\,\AA) and exhibits extra blanketing in the blue (mostly because of the greater mass but identical power).}
\label{fig_he4_he8}
\end{center}
\end{figure*}

Other models for interacting SNe have been produced by \citet{groh14} with \cmfgen, but in this case use a radiative-transfer solver adapted for the conditions of optically-thick stellar winds. In this ``stellar wind'' approach, one assumes a steady-state configuration with a prescribed luminosity at a diffusing inner boundary layer, an optically thick medium at that inner boundary, and an overlying steady flow that reprocesses the impinging continuum flux. However, these assumptions are not well suited for SNe~Ibn at most epochs. As discussed by D22, the physical conditions post-maximum brightness in SNe~Ibn suggest an optically thin nebula cooling through a forest of Fe\two\ lines and a negligible contribution from continuum processes apart from the far-UV range. The observations also suggest the presence of low-velocity dense material (probably in a dense shell formed from the interaction between ejecta and CSM) and powered by a shock (D22). The power originates from within the interaction region rather than being exterior to it. This interaction is also marginally optically thin and thus a non-LTE treatment is preferred. Under these assumptions, the radiative-transfer calculations with \cmfgen\ reproduce satisfactorily the observed spectra of SNe~Ibn (D22). 

Because of the spectral similarity between SN~2020nxt and the SNe~Ibn discussed by D22, we use similar He-rich models. The presence of He\one\ lines at all times suggests a He-rich progenitor.  \citet{woosley_he_19} shows that the persistent presence of He\one\ lines in SNe~Ibn like SN\,2020nxt excludes high-mass progenitors, instead corresponding to a He-star model with an initial mass on the He zero-age main sequence (ZAMS) of $\lesssim 4$\,\msun . For reference, Fig.~\ref{fig_he4_he8} shows the UV and NIR spectra of models based on the composition of a He star explosion from a progenitor that had 4 or 8\,\msun\ initially on the He ZAMS, i.e. the he4 and he8 models. All the material from this He-star model is then placed within a narrow, dense shell, with a profile given by a Gaussian centred on a velocity of 2000\,\kms\ and a characteristic width of 70\,\kms. While the persistent presence of He\one\ lines in SNe~Ibn like SN\,2020nxt excludes high-mass progenitors such as model he8, any model with about 50\% He mass fraction seems suitable. In the He-star models of \cite{woosley_he_19}, this suggests that models he3 (i.e. a He star progenitor of 3\,\msun\ initially on the He ZAMS) to he4 are adequate; we choose the he4 model for the present study. This model is characterised by a total mass of 1.62\,\msun, with 0.92\,\msun\ of He, 0.31\,\msun\ of O, 0.03\,\msun\ of Mg, 0.0014\msun\ of Ca, and a solar metallicity. In our simulations, we assume all the power comes from interaction and thus ignore any radioactive-decay heating.

\subsection {Modeling the Spectra}

\begin{figure*}
\begin{center}
\includegraphics[width=0.8\hsize]{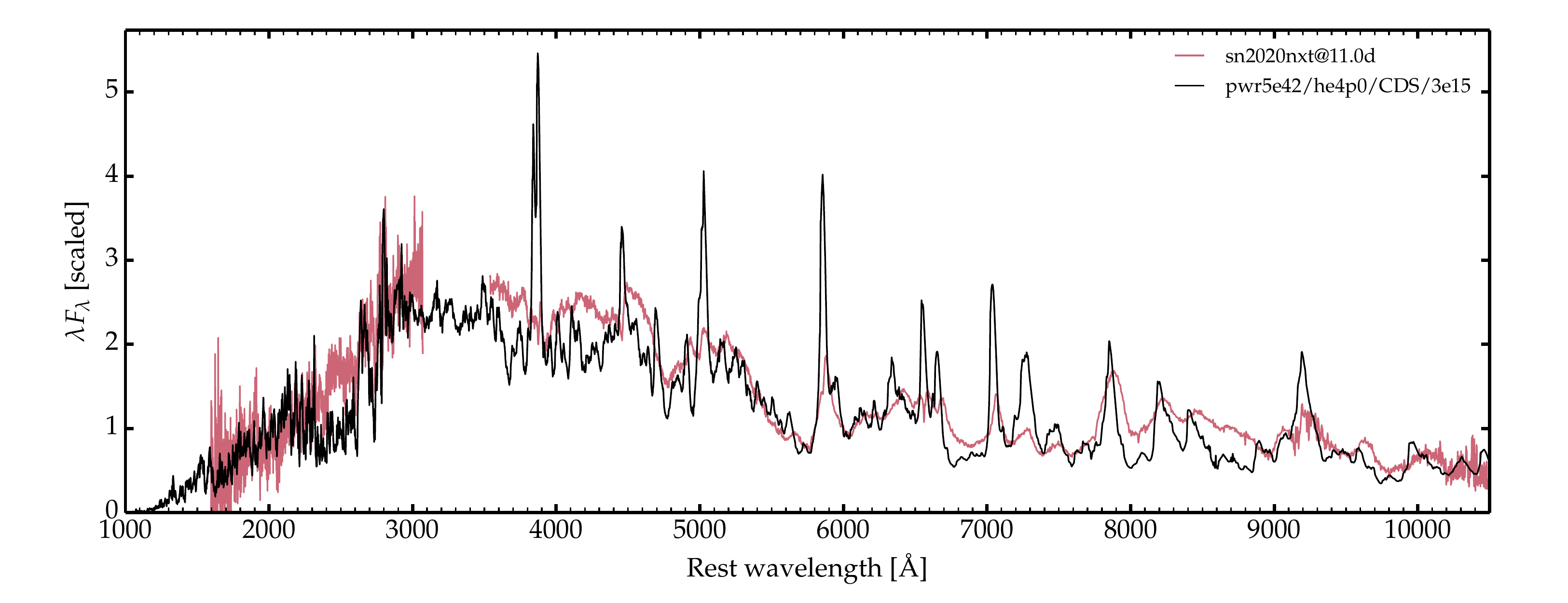}
\includegraphics[width=0.8\hsize]{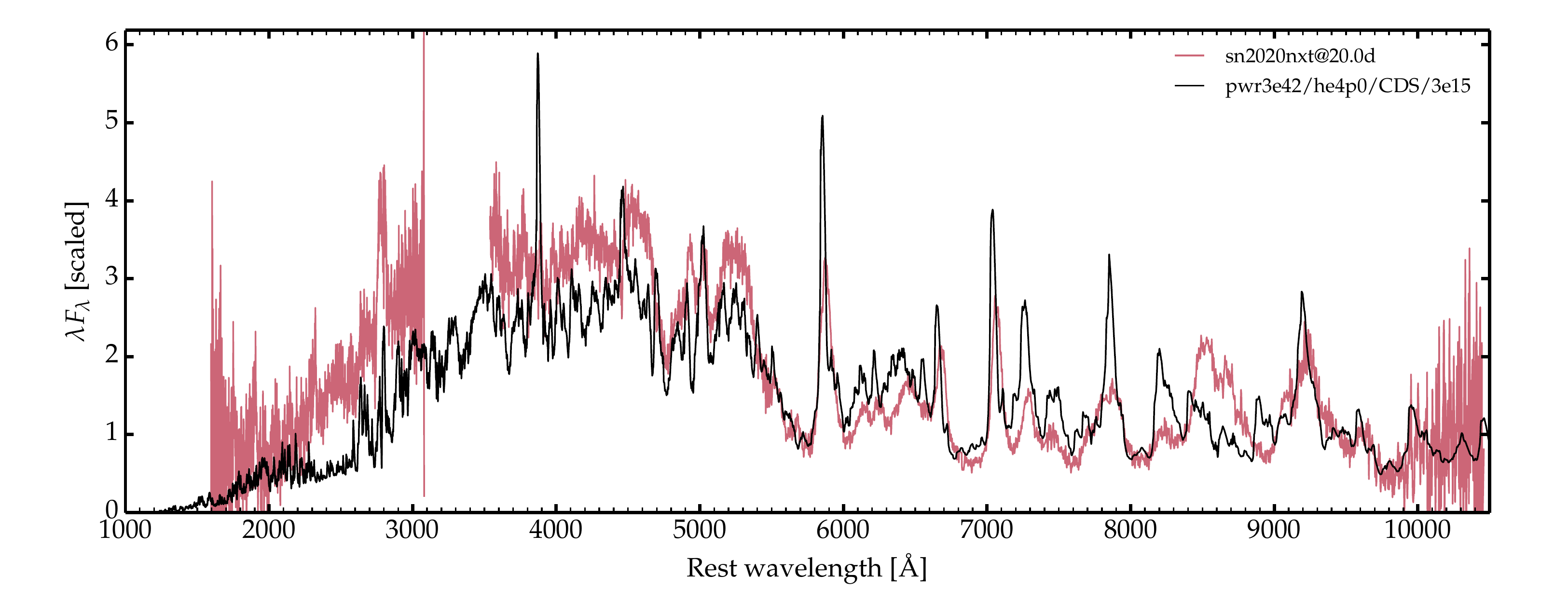}
\includegraphics[width=0.8\hsize]{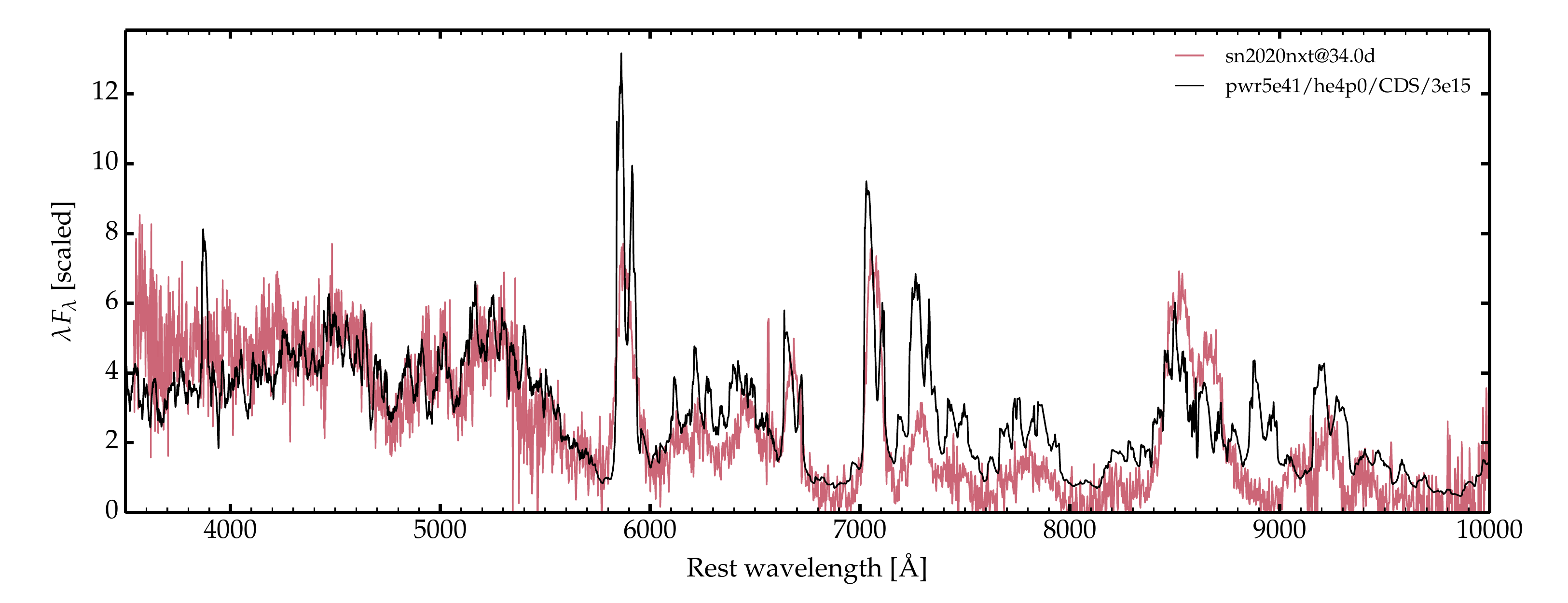}
\includegraphics[width=0.8\hsize]{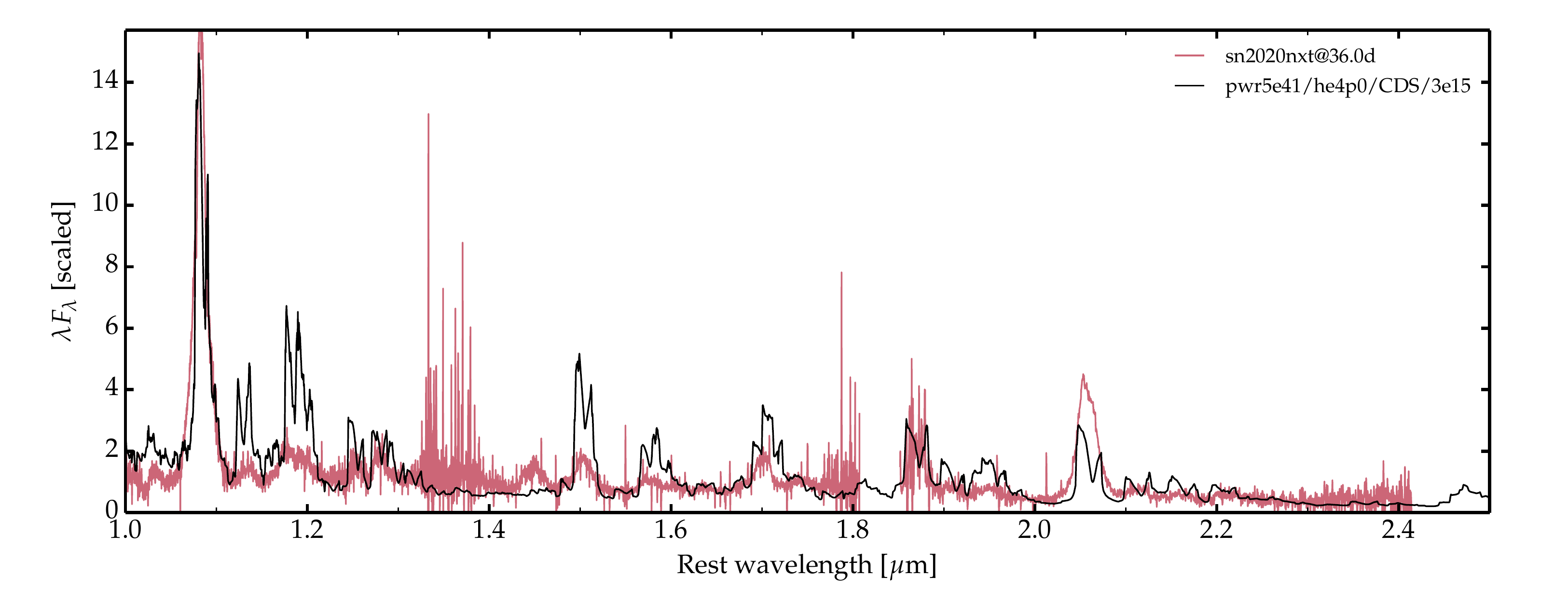}
\caption[]{Comparison of he4 models and multiwavelength spectra of SN~2020nxt at 11, 20, 34, and 36\,days after $o$-band maximum brightness. The top two panels combine {\it HST}/UV and optical data calibrated to contemporarneous photometry.}
\label{fig_mod_obs}
\end{center}
\end{figure*}

\begin{figure*}
\begin{center}
\includegraphics[width=\hsize]{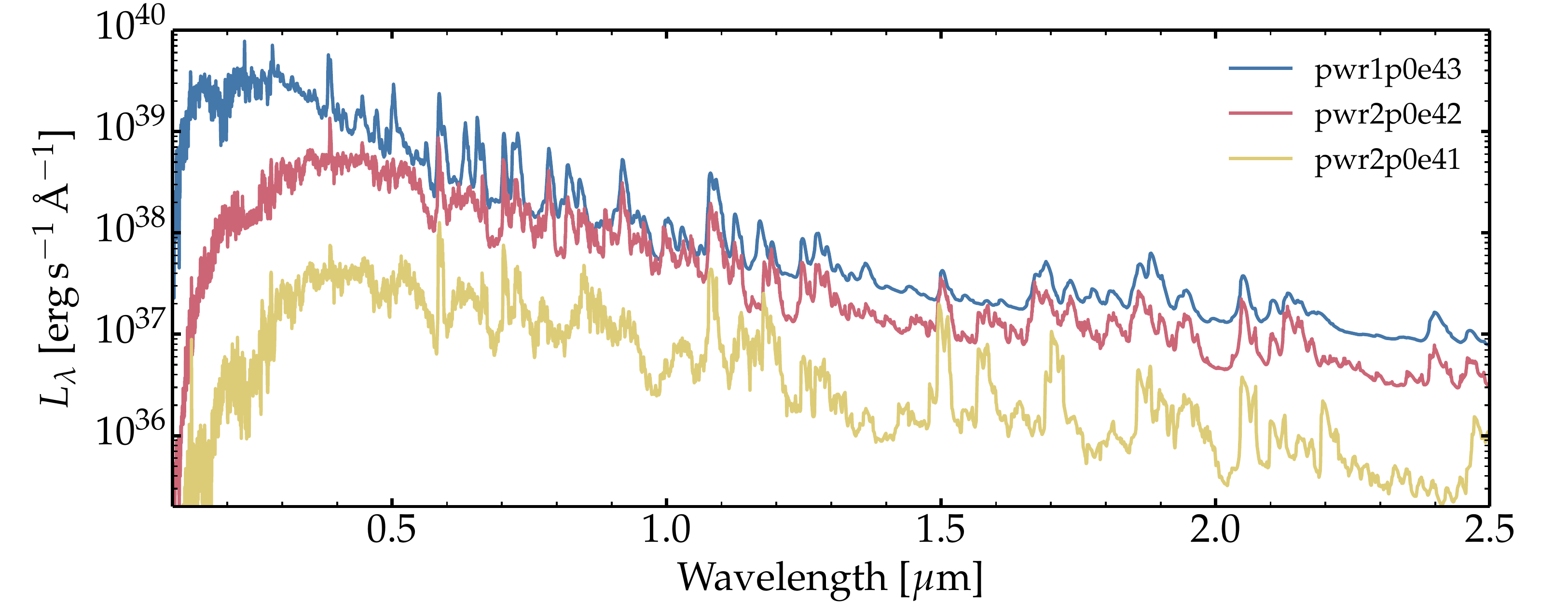}
\caption[]{Grid of interaction calculations based on the he4p0 model, a radius of $3 \times 10^{15}$\,cm, and a velocity of 2000\,\kms\ with powers of $2 \times 10^{41}$, $2 \times 10^{42}$, and $10^{43}$\,erg/s. With decreasing power, the ionisation of the dense shell gets lower and continuum gets weaker especially in the UV region.}
\end{center}
\end{figure*}

Fig.~\ref{fig_mod_obs} shows a comparison between the he4 model and the spectra of SN~2020nxt at multiple epochs after bolometric maximum brightness. For simplicity, we use the same model for all epochs although variations in composition would reduce the discrepancy for certain lines. With increasing time, the only parameter that is changed is power, dropping from $6 \times 10^{42}$ to $5 \times 10^{41}$\,\ergs\ between 10 and 34--36\,days after $o$-band maximum, while the dense shell is set at $3 \times 10^{15}$\,cm and moves at 2000\,\kms. These powers are roughly compatible with the inferred bolometric light curve (Fig.~\ref{fig:bololc}). The radius is not firmly established, but in the similar-looking SN~Ibn 2006jc a pre-SN outburst occurred 2\,yr before the main explosion. So, a representative year-long timescale with line widths suggestive of a velocity of order 1000\,\kms\ yields a representative radius of $3 \times 10^{15}$\,cm.

Given the simplifications of the approach (i.e., the dense shell, the power source, spherical symmetry, etc.), the observations are satisfactorily reproduced. In particular, the landmarks of SNe~Ibn with strong Fe\two\ line emission everywhere below 5500\,\AA\ and the myriad of He\one\ lines is well reproduced at all epochs. Fig.~\ref{fig:line-evolution} further shows the evolution of he ratio of the flux associated with Fe\two\ line emission below 5500\,\AA\ and some individual He\one\ lines. At early times after maximum brightness, our assumption of a narrow dense shell is probably the least adequate and may explain in part the overestimation of the He\one\ line strengths (top panel of Fig.~\ref{fig_mod_obs}). At that time, the flux between lines does not drop to very small values as observed at later epochs and the contrast in flux between the blue and red parts of the optical is moderate. Our model successfully reproduce this feature by employing a greater power ($5 \times 10^{42}$\,erg\,s$^{-1}$). Hence, relative to later epochs when the luminosity is smaller, the stronger continuum flux is caused by the greater ionisation, which also causes a greater ejecta optical depth (the total Rosseland-mean optical depth $\tau_{\rm Ross} = 2.4$). 
  
At subsequent epochs, we adopt the bolometric luminosity inferred in Section~\ref{sec:UVOIRimaging} and the power is thus reduced, first to $3 \times 10^{42}$ (20\,days post-maximum) and then $5 \times 10^{41}$\,erg\,s$^{-1}$ (34--36\,days post-maximum). With a lower power, the dense shell has a lower ionisation, a smaller optical depth ($\tau_{\rm Ross}$ is then 1.4 and 0.55, respectively), and Ca$^+$ eventually dominates over Ca$^{2+}$, causing the strengthening of the Ca\two\ NIR triplet. This feature, never reproduced in the D22 models, may now be explained exclusively through an ionisation (rather than an abundance) effect. In these models, the Mg\two\ lines at 7896, 8234, and 9218\,\AA\ tend to be overestimated at all epochs, although we do reproduce their observed weakening with time --- the strength of these Mg\two\ lines varies between observed SNe Ibn and they eventually weaken, probably as a result of the reduction in ionisation and density (which are also inferred from the strengthening of the Ca\two\ NIR triplet). The change in these two sets of lines can be more easily seen in Fig.~\ref{fig:line-evolution}. The NIR spectrum shown in the bottom panel of Fig.~\ref{fig_mod_obs} yields a satisfactory match to the observations of SN~2020nxt at 35.7\,days, although the model tends to overestimate the strength of some metal lines, in particular those associated with C\one, Mg\one, and Mg\two\ (the same discrepancy affects Mg\two\ lines in optical spectra).

There are two important discrepancies here. First, the model spectra for the first two epochs struggle to reproduce the whole UV and optical range, as well as account for the simultaneous presence of Mg\two\ and Ca\two\ lines. A possible solution to this discrepancy is that the spectrum forms in a more complex environment than adopted here, one that includes a broader range in density and ionisation, as might be expected in a 3D interaction model. Evidence for this comes from another discrepancy in line-profile shapes. As discussed by D22 (in particular their Section~4.2 and Fig.~6), our 1D dense-shell models exhibit a dip as well as a blue-red asymmetry in essentially all strong lines (i.e., He\one\ lines but also those of Mg\two). This arises from a moderate continuum optical-depth effect (the blue-red asymmetry) as well as a line optical depth effect quenching the emission from the regions in the dense shell located in the midplane (at line-of-sight velocities near zero). This feature occurs in many similar dense-shell models and affects H\one\ lines in H-rich models. However, this feature is not observed at similar epochs, even in observations of SNe~Ibn at high resolution and high S/N (e.g., SN~2006jc; \citealt{foley07}). One way of reducing such optical-depth effects in the model is to break up the dense shell into radially- and laterally-confined clumps. The 2D broken-shell models of clumped WR winds carried out by \citet{flores_shell_22} indicate that clumping can resolve this feature. Conversely, the lack of blue-red asymmetry and central dip in He\one\ line profiles of observed SNe~Ibn suggests that the dense shell that forms in those interactions is significantly clumped.

\subsection{Composition and Unique Spectral Line Diagnostics}
\label{sec:linediag}
In numerous SNe~IIn, the thermalised shock radiation within the dense shell at the interface between ejecta and CSM escapes predominantly in the UV, where a multitude of resonance lines of ions with different ionisation potentials provide constraints on the abundances of both iron-group and intermediate-mass elements \citep[e.g.,][]{fransson05,groh14}. Furthermore, the composition alters the colour through the changes in metal line blanketing and plasma cooling processes. The line-profile morphology also constrains the dynamics and geometry \citep[e.g.,][]{dessart11}. In the context of SN~2020nxt, the NUV range does not reveal such a variety of lines because of the much cooler gas and its low ionisation at the epochs observed. Despite its low abundance, Fe (essentially at solar metallicity) represents the strongest coolant for the gas. This is in stark contrast from standard SN ejecta at nebular times where the low gas density favours the formation of strong forbidden lines that dominate the cooling. Here, because of the material compression associated with the interaction, such lines do not form. At earlier times, when a fraction of the CSM has not been shocked and compressed but is instead lower-density slow gas influenced by the radiation injected at the shock, the SN may emit a much bluer spectrum with lines from ions with a high ionisation potential.

\begin{figure*} 
\centering
\includegraphics[width=\textwidth]{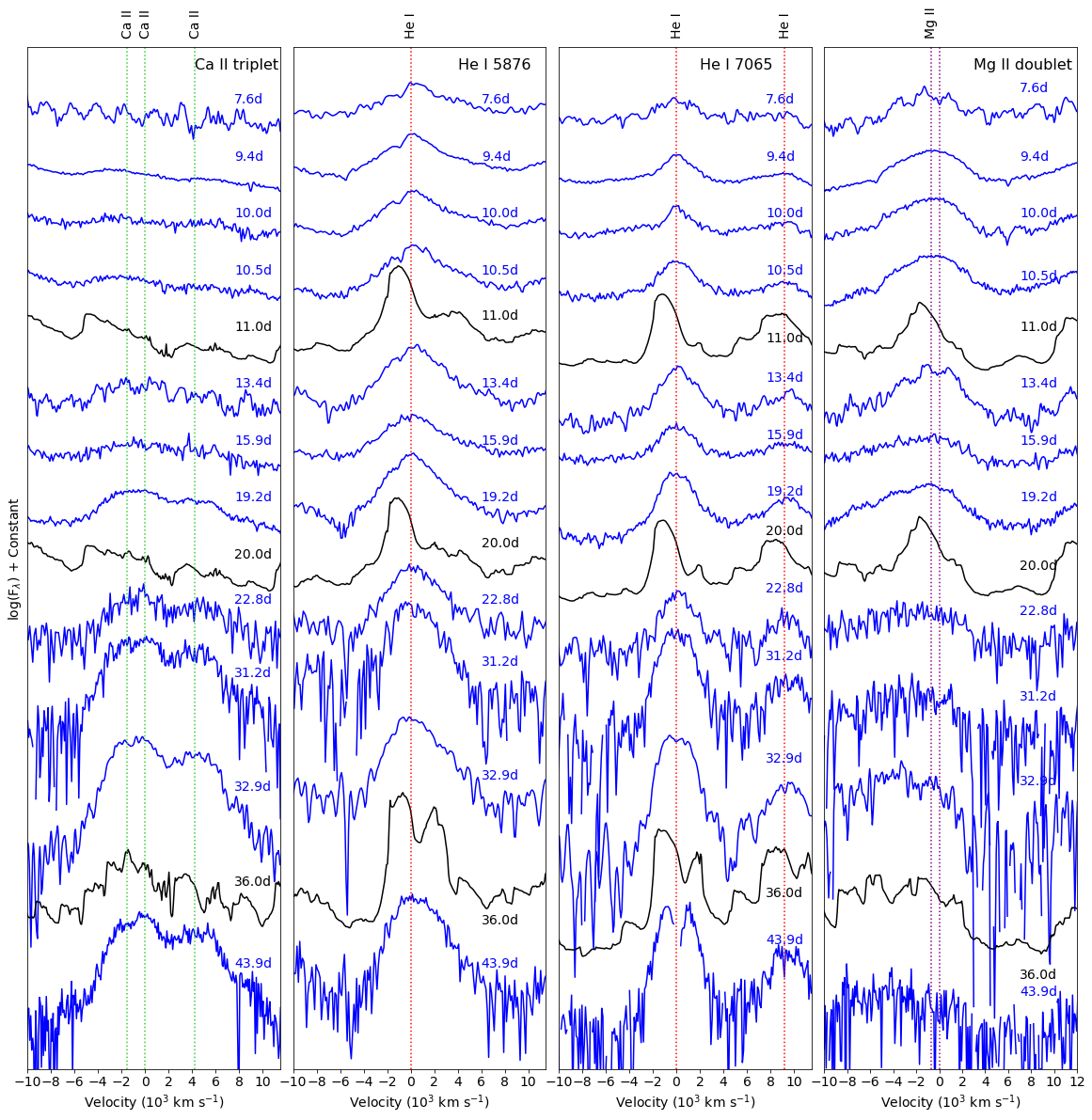}
    \caption{The profiles of He\one\ 5876, 7065\,\AA\ , Mg\two\ doublet around 7891\,\AA\ and Ca\two\ triplet around 8542\,\AA\ from observed (blue) and theoretical (black) spectra, plotted in velocity space. Spectra are normalised to the 6100--6300\,\AA\ continuum. Prominent features are marked with vertical dashed lines. Phases are labeled near each spectrum.
    }
    \label{fig:lineVel}
\end{figure*}

\begin{figure*}
    \centering
    \includegraphics[width=0.49\textwidth]{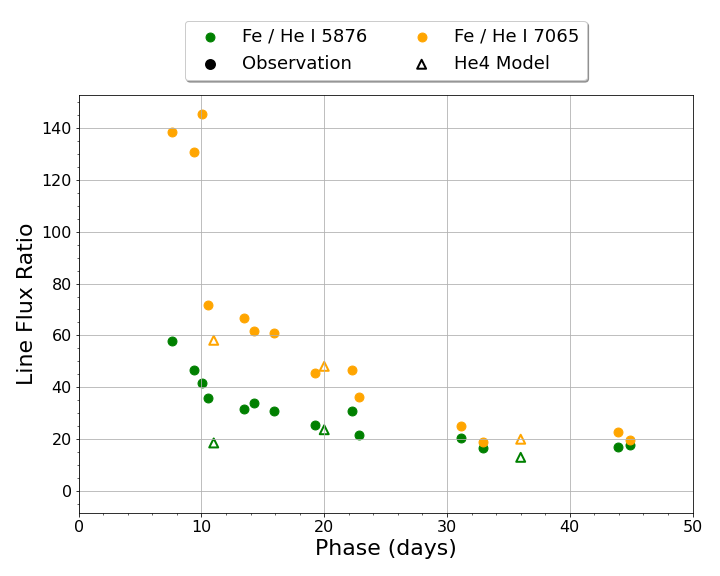}
    \includegraphics[width=0.49\textwidth]{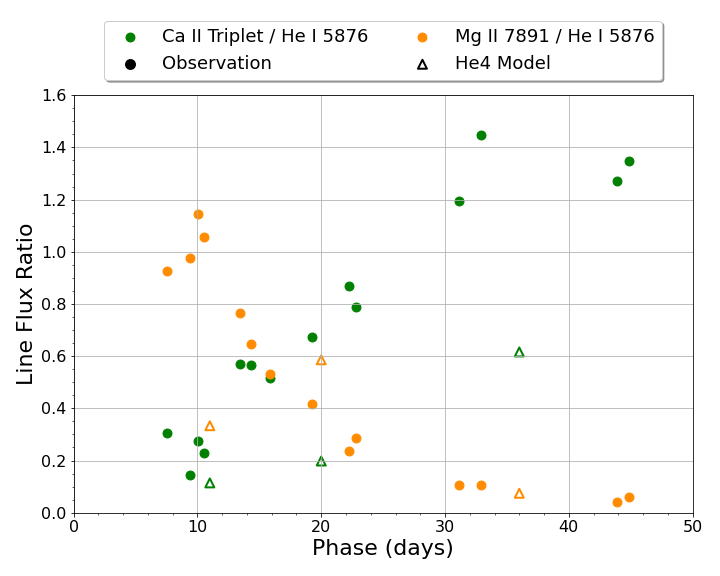}
    \caption{\textbf{Left:} The evolution of the line flux ratio between Fe\two\ continuum below $5500\AA$ and the prominent He\one\ 5876\,\AA\ . \textbf{Right:} The evolution of the line flux ratio between Mg\two\ 7891\,\AA, the Ca\two\ triplet around 8542\,\AA and the prominent He\one\ 5876\,\AA\ . The phases are relative to the $o$ band peak. 
    }
    \label{fig:line-evolution}
\end{figure*}

The flux of a few dominant emission lines has been measured (after subtracting the continuum) for line diagnostics. For the Mg\two\ doublet around 7891\,\AA\ and the Ca\two\ triplet around 8542\,\AA\ that largely overlap, the flux in the whole line region is taken into account. We integrate flux in the range 4000--5500\,\AA\ to approximate the total flux of Fe\two\ emission lines. Owing to its complicated and extended profile, the Fe\two\  emission continuum is indistinguishable from the underlying blackbody continuum, so the result is unavoidably an overestimation of the real flux of Fe\two\ emission. However, beyond $\sim 15$\,days after peak, the emission lines gradually dominate the spectra as the continuum fades, so the total optical flux below 5500\,\AA\ can serve as a good approximation of the Fe line emission at later phases. A few relatively weak lines, such as He\one\ 4471, may also contribute to the systematic bias in measuring Fe\two\ emission. 
The left panel of Fig.~\ref{fig:line-evolution} shows the ratios between the flux intensity of the continuum dominated by Fe\two\ emission lines below 5500\,\AA\ and He\one\ 5876, 7065\,\AA\ at different epochs, and the right panel shows the evolution of the flux ratio between He\one\ 5876\,\AA\ and the Mg\two\ doublet or the Ca\two\ triplet. We also include the ratio measured from the simulation as discussed in Section~\ref{sec:analysis} following the same line-measurement scheme.

The [Fe/He\one] ratio decreases with time as the underlying continuum weakens and He\one\ features become stronger, gradually levelling off after $\sim 20$\,days with ratios still $\gtrsim 20$. This feature shows the dominance of Fe \two\ emission as the major coolant throughout all phases. Other trends include the fading of the Mg\two\ doublets and the increase in the Ca\two\ flux. After $\sim 30$\,days, there is no obvious evolution in any of these line ratios. 

In general, the model predictions agree with observations quantitatively except for a few systematic biases. At $\sim 10$\,days after peak brightness, all of the line ratios in Figure~\ref{fig:line-evolution} are underestimated, likely caused by the fact that He\one\ is largely overestimated at early phases. As discussed in Section~\ref{sec:analysis}, such a systematic bias is likely caused by our oversimplified assumption of a narrow dense shell in the modelling. Another issue is that the model fails to reproduce the Ca\two\ triplet, as can be seen from the first panel of Figure~\ref{fig:lineVel}. Such an effect can now be explained by an ionisation effect, as discussed in Section~\ref{sec:analysis}. Overall, the model predictions show an excellent match with the observations, given the oversimplified assumptions used.

\subsection {Model Implications and Discussion}
\label{sec:discussion}


The best-fitting model for SN 2020nxt is a moderate-mass He star powered by interaction. 
We show some good matches for a He-star model of 4\,\msun\ initially on the He ZAMS, which corresponds to a pre-SN star mass of 3.16\,\msun\ and a mass of 18.11\,\msun\ on the H ZAMS. We have not covered the full parameter space of composition, powers, and locations for the shell, but tests have shown that the persistence of He\one\ lines at all epochs requires a large He mass fraction relative to the total mass. Given the modest mass of the He-rich shell (i.e, the He/C and the He/N shells) of $\sim 1$\,\msun\ in massive stars at the time of core collapse (D22, Fig.~1), this constraint suggests that the total pre-SN mass cannot be much greater than 3--4\,\msun. This effectively excludes massive stars with  $M>$18\,\msun\ on the H ZAMS such as WR stars and points instead to lower-mass massive stars that, in addition, have evolved in an interacting binary and lost their H-rich envelope through mass transfer from even lower mass stars. This type of binary interaction is required, since red supergiant winds are too weak to remove the H envelope in this initial mass range \citep{beasor20,beasor21}.  It also opens the possibility that some SNe~Ibn arise instead from low-mass stars such as He white dwarfs. 

Connecting SNe~Ibn to SNe~Icn, FBOTs, and more generally to fast-evolving luminous transients is straightforward, as demonstrated by D22. For H-free ejecta, the mass of CSM that one expected 
from a massive star tends to be smaller than typically encountered in H-rich mass stars. Furthermore, metal-rich (or H-free) ejecta are poor donors of free electrons and thus have a lower electron-scattering opacity (the associated mass-absorption coefficient is reduced by a factor of a few to a few tens). Consequently, the CSM has a lower optical depth and traps radiation for a shorter period, naturally leading to shorter rise times. If one assumes a similar energy injection from kinetic energy of the inner ejecta in the interaction scenario, this implies a greater peak luminosity. For example, radiating at $10^{43}$\,erg\,s$^{-1}$ for 100\,days is energetically identical to radiating $10^{44}$\,erg\,s$^{-1}$ for 10\,days or $10^{45}$\,erg\,s$^{-1}$ for 1\,day. The critical energy constraint is not the peak luminosity but the time integral of the bolometric luminosity. One may thus expect a natural trend for greater luminosity for shorter-lived transients (see Section~3 and Fig.~3 of D22). 



\section{Conclusion}\label{section:conclusion}

We present multiwavelength observations of the Type Ibn SN~2020nxt. In addition, we provide a comparison of its spectroscopic evolution to the state-of-the-art radiative transfer simulations based on D22, supporting a He-star progenitor of 4\,M$_\odot$ initially on the He ZAMS that lost its $\sim 1$\,M$_\odot$ He-rich envelope in the years prior to explosion. 
There are discrepancies in reproducing the UV flux, some line ratios, and the line profiles, but there is hope in solving these issues with a more realistic, 3D interaction model of a clumpy dense shell.. From the analysis of the pre-SN light curve in the \atlas\ $o$ band within 3\,yr of the SN discovery, no signature of a pre-SN eruption has been found with a limiting magnitude of $M_o\lesssim-14.8$mag. The relatively low mass of the progenitor star disfavours the single massive star scenario, indicating that the progenitor system is likely to be an interacting binary system in which CSM may arise from the mass-transfer process. This result is consistent with low mass progenitor star models for three other SNe Ibn fit by D22: SNe 2006jc, 2011hw, and 2018bcc.

UV spectra were important in this work. 
In addition to the spectroscopic evolution, we show that a fundamental quantity of interest is the fraction of the total kinetic energy that has been extracted from the impacting shell. This fractional kinetic energy is radiated, typically on a diffusion timescale, and produces the bolometric light curve of the SN. Because the emitting interaction region is often hot and ionised, a significant fraction of that radiation falls in the UV range and can be critical for differentiating between progenitor models (i.e., Figure \ref{fig_he4_he8}). The UV spectra were also key to securing the identification of Fe\two\ line forest as the main contributor to the emission below 5500\,\AA, emphasising that this Fe\two\ emission actually extends down to 1000\,\AA\ where it vanishes. SNe~Ibn are known for this strong emission in the blue part of the optical, and it is clearly not related to continuum emission; if that had been the case, the flux would have been much stronger in the UV.

Although very challenging, earlier UV will better help constrain the progenitors of SNe~Ibn and FBOTs. At earlier phases (i.e., around or before bolometric maximum), the underlying continuum would have been stronger in the UV and the ionisation likely would have been higher, at a time when the optical spectrum tends to be blue and featureless (e.g., like in FBOTs). It is at such times that lines for He\two, C\three, C\four\ or N\five\ could be seen in the UV, at a time when the optical spectrum tends to be blue and featureless (e.g., like in FBOTs).

In summary, these results provide an important perspective on the broader conversation about the evolutionary pathways of stripped-envelope SNe. Like non-interacting SNe Ib/c, the progenitor star likely went through multiple eruptions or interaction process to lose its H envelope and form massive He rich CSM around it (though such activities were not detected in the pre-SN \atlas\ light curve). At the same time, the H-free CSM around such SNe can naturally produce other fast luminous transients like SNe~Icn or FBOTs that are featureless in optical but full of high ionization features in UV at early time. Thus, the early UV spectra will be a key to understand the physical nature of SNe~Ibn, fast transients, and all stripped-envelope SNe in the future.

\section*{Acknowledgement}

Funding for this program was provided by NASA/{\it HST} grant AR-14295 from        
the Space Telescope Science Institute (STScI), which is operated by the Association of Universities for Research in Astronomy (AURA), Inc., under NASA contract NAS5-26555.
T.S. has been supported by the J\'anos Bolyai Research Scholarship of the Hungarian Academy of Sciences, as well as by the FK134432 grant of the National Research, Development and Innovation Office of Hungary and the \'UNKP 22-5 New National Excellence Programs of the Ministry for Culture and Innovation from the source of the National Research, Development and Innovation Fund. A.V.F.'s supernova group at UC Berkeley has been supported by the Christopher R. Redlich Fund, Frank and Kathleen Wood, Alan Eustace, and numerous other individual donors. P.K. is supported by NSF grant AST-1908823. D. A. Coulter acknowledges support from the National Science Foundation Graduate Research Fellowship under Grant DGE1339067.
C.G. is supported by a VILLUM FONDEN Young Investigator Grant (project number 25501).

The data presented here were obtained in part with ALFOSC, which is provided by the Instituto de Astrofisica de Andalucia (IAA) under a joint agreement with the University of Copenhagen and NOT.
This work makes use of data taken with the Las Cumbres Observatory global telescope network.  The LCO group is funded by NSF grants AST-1911151 and AST-1911225.
Some of the data presented herein were obtained at the W. M. Keck
Observatory, which is operated as a scientific partnership among the
California Institute of Technology, the University of California, and
NASA; the observatory was made possible by the generous financial
support of the W. M. Keck Foundation.

YSE-PZ was developed by the UC Santa Cruz Transients Team with support from The UCSC team is supported in part by NASA grants NNG17PX03C, 80NSSC19K1386, and 80NSSC20K0953; NSF grants AST-1518052, AST-1815935, and AST-1911206; the Gordon \& Betty Moore Foundation; the Heising-Simons Foundation; a fellowship from the David and Lucile Packard Foundation to R. J. Foley; Gordon and Betty Moore Foundation postdoctoral fellowships and a NASA Einstein fellowship, as administered through the NASA Hubble Fellowship program and grant HST-HF2-51462.001, to D. O. Jones; and a National Science Foundation Graduate Research Fellowship, administered through grant No. DGE-1339067, to D. A. Coulter.

KAIT and its ongoing operation at Lick Observatory were made possible by donations from Sun Microsystems, Inc., the Hewlett-Packard Company, AutoScope Corporation, Lick Observatory, the U.S. NSF, the University of California, the Sylvia \& Jim Katzman Foundation, and the TABASGO Foundation. A major upgrade of the Kast spectrograph on the Shane 3\,m telescope at Lick Observatory was made possible through generous gifts from William and Marina Kast as well as the Heising-Simons Foundation.
The authors acknowledge the help of James Sunseri, Matt Chu, Michael May, Nachiket Girish, Raphael Baer-Way, Teagan Chapman, Andrew Hoffman, and Asia deGraw in obtaining Lick Nickel 1\,m data.
We appreciate the excellent assistance of the staff at Lick Observatory.         
Research at Lick Observatory is partially supported by a generous gift from Google.\\

"The Young Supernova Experiment (YSE) and its research infrastructure is supported by the European Research Council under the European Union's Horizon 2020 research and innovation programme (ERC Grant Agreement 101002652, PI K.\ Mandel), the Heising-Simons Foundation (2018-0913, PI R.\ Foley; 2018-0911, PI R.\ Margutti), NASA (NNG17PX03C, PI R.\ Foley), NSF (AST-1720756, AST-1815935, PI R.\ Foley; AST-1909796, AST-1944985, PI R.\ Margutti), the David \& Lucille Packard Foundation (PI R.\ Foley), VILLUM FONDEN (project 16599, PI J.\ Hjorth), and the Center for AstroPhysical Surveys (CAPS) at the National Center for Supercomputing Applications (NCSA) and the University of Illinois Urbana-Champaign.

The UCSC team is supported in part by NASA grant NNG17PX03C, NSF grant AST--1815935, the Gordon \& Betty Moore Foundation, the Heising-Simons Foundation, and by a fellowship from the David and Lucile Packard Foundation to R.J.F.


A subset of the data presented herein were obtained at the W.\ M.\ Keck Observatory. NASA Keck time is administered by the NASA Exoplanet Science Institute. Data presented herein were obtained at the W. M. Keck Observatory from telescope time allocated to the National Aeronautics and Space Administration through the agency’s scientific partnership with the California Institute of Technology and the University of California. The Observatory was made possible by the generous financial support of the W. M. Keck Foundation. The authors wish to recognise and acknowledge the very significant cultural role and reverence that the summit of Maunakea has always had within the indigenous Hawaiian community.  We are most fortunate to have the opportunity to conduct observations from this mountain.

Pan-STARRS is a project of the Institute for Astronomy of the University of Hawaii, and is supported by the NASA SSO Near Earth Observation Program under grants 80NSSC18K0971, NNX14AM74G, NNX12AR65G, NNX13AQ47G, NNX08AR22G, 80NSSC21K1572 and by the State of Hawaii.  The Pan-STARRS1 Surveys (PS1) and the PS1 public science archive have been made possible through contributions by the Institute for Astronomy, the University of Hawaii, the Pan-STARRS Project Office, the Max-Planck Society and its participating institutes, the Max Planck Institute for Astronomy, Heidelberg and the Max Planck Institute for Extraterrestrial Physics, Garching, The Johns Hopkins University, Durham University, the University of Edinburgh, the Queen's University Belfast, the Harvard-Smithsonian Center for Astrophysics, the Las Cumbres Observatory Global Telescope Network Incorporated, the National Central University of Taiwan, STScI, NASA under grant NNX08AR22G issued through the Planetary Science Division of the NASA Science Mission Directorate, NSF grant AST-1238877, the University of Maryland, Eotvos Lorand University (ELTE), the Los Alamos National Laboratory, and the Gordon and Betty Moore Foundation."


\noindent{\bf Data availability}\\

The data underlying this article will be shared on reasonable request to the corresponding author.\\

\bibliographystyle{mnras}
\bibliography{references}

\end{document}